\documentclass[11pt]{article}
\usepackage{amsmath}
\usepackage{graphicx}
\usepackage{enumerate}
\usepackage{amsfonts}
\usepackage{url} % not crucial - just used below for the URL 

\usepackage[margin=10mm]{caption}
\usepackage{geometry}
\usepackage{mathptmx}
\usepackage{graphics,epsfig,latexsym,amssymb,amsbsy,amsmath}
\usepackage{amsthm}
\usepackage[margin=15mm]{caption}

\renewcommand{\baselinestretch}{1.2}
\newcommand{\bX}{{\bf X}}

\newcommand{\bA}{{\bf A}}
\newcommand{\bD}{{\bf D}}
\newcommand{\bx}{{\bf x}}
\newcommand{\btheta}{\mbox{\protect\boldmath $\theta$}}
\newcommand{\bgamma}{\mbox{\protect\boldmath $\gamma$}}
\newcommand{\etab}{\mbox{\protect\boldmath $\eta$}}
\newcommand{\bmu}{\mbox{\protect\boldmath $\mu$}}
\newcommand{\bbeta}{\mbox{\protect\boldmath $\beta$}}
\newcommand{\bLambda}{\mbox{\protect\boldmath $\Lambda$}}

\newcommand{\E}{\mathbb{E}}

\newcommand{\BFI}{{\mbox{\footnotesize{BFI}}}}
\newcommand{\WAV}{{\mbox{\footnotesize{WAV}}}}
\newcommand{\single}{{\mbox{\footnotesize{single}}}}
\newcommand{\com}{{\mbox{\footnotesize{com}}}}
\newcommand{\MSER}{\mbox{MSE}}
\newcommand{\MSET}{\mbox{MSET}}

\begin{document}

\def\spacingset#1{\renewcommand{\baselinestretch}%
{#1}\small\normalsize} \spacingset{1.2}

{
  \title{\bf Bayesian Federated Inference for regression models based on non-shared multicenter data sets from heterogeneous populations}
  \author{Marianne A Jonker \thanks{corresponding author: marianne.jonker@radboudumc.nl}\hspace{.2cm}\\
    Research Institute for Medical Innovation, Science department IQ Health, Section Biostatistics, \\
    Radboudumc, Nijmegen, The Netherlands\\
    and \\
    Hassan Pazira \\
    Research Institute for Medical Innovation, Science department IQ Health, Section Biostatistics, \\
    Radboudumc, Nijmegen, The Netherlands\\
    and \\
Anthony CC Coolen\\
    Donders Institute, Faculty of Science, Radboud University, 
    Nijmegen, The Netherlands\\
    Saddle Point Science Europe, Mercator Science Park, 
    Nijmegen, The Netherlands
    } 
}  

\maketitle

\abstract{To estimate accurately the parameters of a regression model, the sample size must be large enough relative to the number of possible predictors for the model. In practice, sufficient data is often lacking, which can lead to overfitting of the model and, as a consequence, unreliable predictions of the outcome of new patients. Pooling data from different data sets collected in different (medical) centers would alleviate this problem, but is often not feasible due to privacy regulation or logistic problems. An alternative route would be to analyze the local data in the centers separately and combine the statistical inference results with the Bayesian Federated Inference (BFI) methodology. The aim of this approach is to compute from the inference results in separate centers what would have been found if the statistical analysis was performed on the combined data. We explain the methodology under homogeneity and heterogeneity across the populations in the separate centers, and give real life examples for better understanding. Excellent performance of the proposed methodology is shown. An R-package to do all the calculations has been developed and is illustrated in this paper. The mathematical details are given in the Appendix.}

\bigskip


%\jnlcitation{\cname{%
%\author{Jonker MA.},
%\author{Pazira H.}, and
%\author{Coolen ACC}.
%\ctitle{???} \cjournal{\it ???} \cvol{???.}}}

\maketitle

\section{Introduction}
Prediction models aim to predict the outcome of interest for individuals (or subjects), based on their values of the covariates in the model. To build a prediction model by selecting covariates and estimating the corresponding regression parameters, the sample size should be sufficiently large. If too many variables (possible covariates) relative to the number of events or observations are included, the model may become overly flexible and erroneously `explain' noise or random variations in the data, rather than estimating meaningful relationships between the covariates and the outcome. This is called overfitting and may lead to unreliable predictions of the outcome for new individuals\cite{Harrell}. To overcome overfitting a minimum of 10 observations or events per variable (EPV) is often advised\cite{Harrell2001, Harrell1984}. Based on this criterion, data sets are often too small to take all available variables in consideration. Merging different data sets from different (medical) centers could in principle alleviate the problem, but is often difficult for regulatory and logistic reasons. An alternative route would be to analyse the local data in the centers and combine the obtained inference results intelligently. With this approach the (individual) data do not need to be shared across centers. In this paper, we focus on methodology to combine the local inference results for estimating  parametric regression models for a general population of interest. The data sets in the centers are considered as samples from this population. 

In literature, several methods have been described. Probably the best-known strategy to obtain effect estimates from different inference results, is meta-analysis\cite{Borenstein}. In a meta-analysis, relevant, already published results are combined. Here we consider the situation where the local analyses have yet to be performed. This means that the collaborating centers discuss in advance which local analyses will be performed and what inference results should be shared to build the final combined model. It also means that more information can be shared than is usually available in publications, like the estimated covariance matrix of the estimators of the model parameters.  

Federated Learning (FL) is a machine learning approach that was developed several years ago, mainly for analyzing data from different mobile devices \cite{McMahan}. It aims to construct from the inference results obtained in the separate centers, what would have been found if the analysis was performed on the combined data set. With this approach, the local data stay at their owners’ centers, only parameter estimates are cycled around and updated based on the local data until a convergence criterion is met. In recent years the FL approach has improved quite a bit (e.g., on optimization in the local centers and the aggregation of the local results, dealing with heterogeneity and client-drift \cite{Li2020, Zhu, Karimireddy, Shi}, methodology for causality related research questions\cite{Vo, Han}). Also FL in a Bayesian setting for deep learning models has been proposed\cite{Maddox, Al-Shedivat,Cao, Chen}. The posterior distributions are estimated in the local centers and communicated to the central server for aggregation. However, practically this Bayesian procedure is challenging, especially for deep learning models due to the high dimensionality of the parameters. An overview of the most important recent developments and a list of references is given in Liu et al.\cite{Liu}. FL performs excellently in e.g., image analysis \cite{Rieke,Sheller,Gafni} or for data from mobile devices, but has clearly some drawbacks in other applications. For instance, apart from obvious ones such as data security and convergence problems, if one aims to estimate statistical models based on inference results from different medical centers, one needs to handle challenges like heterogeneity of the populations across centers, clustering of centers, center-specific covariates (like location), missing covariates in the data, and the fact that data may be stored in different ways (covariates are named differently or are even defined differently). Furthermore, most FL strategies require many iterative inference cycles across the local centers. In case the centers are hospitals (the situation we are considering here), a cycling mechanism is complex and may lead to considerable extra work; a one-shot approach is preferred.

Also in the field of distributed statistical inference, multiple strategies have been proposed to combine inference results from different computers (centers)\cite{Gaoa}. To cope with massive data sets which can not be analyzed on a single computer, a data set is divided into smaller data sets, which are analyzed separately and the results are combined afterwards. An interesting one-shot algorithm has been proposed by Jordan et al.\cite{Jordan}. They proposed a communication-efficient surrogate likelihood framework for distributed statistical inference for homogeneous data. Instead of maximizing the full likelihood for regular parametric models or the penalized likelihood in high-dimensional models, this surrogate likelihood is maximized. The surrogate likelihood expression was determined so that only a minimum amount of information is transferred from the local machines to the central server (of the order $O(d)$ bits where $d$ is the dimension of the parameter space). Later, the method was generalized to be able to deal with certain forms of heterogeneity\cite{Duan2022}.  

In this paper we describe the BFI framework for parametric regression models. This methodology was developed especially for combining inference results from different centers to estimate statistical (regression) models without the need for repeated communication rounds with the local centers. In every center the data are analysed only once and the inference results (parameter and accuracy estimates) are sent to a central server, where the local inference results are combined. Explicit expressions for the combined (BFI) estimators in terms of the local inference results have been derived. Via these expressions the BFI estimates can be easily updated at a later moment if the data collection or the analysis in several centers are delayed, without contacting all other centers again (this would not be possible when using an iterative updating mechanism). The fact that only one communication round is sufficient is important in our (medical) setting, since  assistance from the local medical and technical staff are needed every time local analyses are performed. 

The BFI estimates are defined as the maximizers of a surrogate expression of the full log posterior density. This expression depends on the local estimates and is different from the one proposed by Jordan (2018)\cite{Jordan}. In the BFI framework more information (of the order $O(d^2)$) is shared with the central server than would normally be acceptable in a FL or distributed statistical inference setting. This additional information improves the accuracy of the estimator. The BFI methodology was developed for estimating (low-dimensional) GLMs. High dimensional models (with large $d$), typically the models of interest in FL and distributed statistical inference, are not the focus of the BFI methodology; estimation accuracy is more important than communication efficiency.        

The mathematical theory of the BFI methodology for parameteric models, like GLMs, was published by the authors in Jonker et al.\cite{Jonker}. In this paper, we extend the theory further to allow for different kinds of heterogeneity between the centers. Among others, we consider the situation in which there is heterogeneity in the population characteristics, there is clustering, the distribution of the outcome variable is shifted, and the regression or nuisance parameters differ between the centers. The asymptotic distributions of the BFI estimators are derived and it is proven that the estimators are asymptotically efficient. Asymptotically, no information is lost if the data from the centers cannot be combined. These asymptotic distributions of the estimators are used for the construction of credible intervals. For finite samples (by means of simulation studies) and asymptotically, the BFI estimators are compared to the estimators that are obtained by averaging the local estimators (weighted for local sample size). In this paper, we also focus on applications: a data example is given and the R code (from our R package BFI \cite{Pazira_package}) for analyzing the data with the BFI methodology is explained. 

This paper is organized as follows. In Section \ref{sec: BFI} the BFI framework for generalized linear models for homogeneous sub-populations in the local centers is explained. In Section \ref{sec: hetero} different types of heterogeneity across these sub-populations and data sets are described and, moreover, it is explained how the BFI methodology can be adjusted to takes these into account. To study the performance of the BFI method in different settings, the results of simulation studies are described in Section \ref{sec: performance}.  In the same section also the analysis of a heterogeneous data set using the BFI methodology is described. A discussion is given in Section \ref{sec: Discussion}. The paper ends with three appendices. In the first appendix we explain how to do the analysis with our R package, the second appendix  contains the mathematical details of the derivation of the estimators  and in the third appendix the asymptotic distributions of the BFI and the weighted average estimators are derived and compared.

\section{The Bayesian Federated Inference (BFI) framework}
\label{sec: BFI}
Suppose that data of $L$ medical centers are locally available, but these data sets cannot be merged to a single integrated data set for statistical analysis. The data for individual $i$ from center $\ell$ is denoted as the pair $(\bx_{\ell i},y_{\ell i})$ with $\bx_{\ell i}$ a vector of covariates and $y_{\ell i}$ the outcome of interest. Let $\bD_\ell$ denote the data subset in center $\ell$: 
\begin{eqnarray}
\bD_\ell=\{(\bx_{\ell 1},y_{\ell 1}),\ldots,(\bx_{\ell n_\ell},y_{\ell n_\ell})\}, \nonumber
\end{eqnarray}
where $n_\ell$ denotes the number of individuals in subset $\ell$, $\ell=1,\ldots,L$, and let $\bD$ be the fictive combined data set (the union of the subsets $\bD_1,\ldots,\bD_L$).

The data pair $(\bx_{\ell i},y_{\ell i})$ is the realisation of the stochastic pair $(\bX_{\ell i},Y_{\ell i})$. Suppose that the variables $(\bX_{\ell i},Y_{\ell i}), i=1,\ldots,n_\ell, \ell=1,\ldots, L$ are independent and identically distributed, and that $\bX_{\ell i}$ and $Y_{\ell i}$ are linked via a generalized linear model (GLM) with link function $h$:
\begin{align*}
h\big(\mathbb{E}(Y_{\ell i}|\bX_{\ell i},\etab,\bbeta)\big) = \bbeta^t \bX_{\ell i},
\end{align*}
where $\bbeta$ is a vector of unknown regression parameters and $\etab$ a vector of unknown nuisance parameters. If the first element in the covariate vector $\bX_{\ell i}$ equals one for all individuals, the model includes an intercept. \footnote{ The theory in this paper holds for any parametric regression model, but for simplicity of notation we focus on GLMs only.}
%For, e.g., the logistic and Poisson regression models the link functions are fully specified and no parameter $\etab$ exists, whereas for the linear regression model the parameter $\etab$ represents the variance of $Y_{\ell i} | \bX_{\ell i}$. 

For $\btheta_1:=(\etab, \bbeta)$, the conditional density of $Y_{\ell i}|(\bX_{\ell i}=\bx,\btheta_1)$ is given by $y|\bx,\btheta_1 \to p(y|\bx,\btheta_1)$ and for the vector of covariates $\bX_{\ell i}| \btheta_2$ this is $\bx|\btheta_2 \to p(\bx|\btheta_2)$, for $\btheta_2$ a parameter vector.\footnote{we use the letter $p$ for any density. From the arguments it is clear which density is actually meant.} Then, for $\btheta:=(\btheta_1,\btheta_2)$ it follows that the density of $y,\bx|\btheta$ can be written as $y,\bx|\btheta \to p(y,\bx|\btheta) \;=\; p(y|\bx,\btheta_1) p(\bx|\btheta_2)$. We work in a Bayesian setting; $\btheta$ is stochastic as well. For mathematical simplicity, we assume statistical independence between $\btheta_1$ and $\btheta_2$. Thus,  $p(\btheta_1,\btheta_2)=p(\btheta_1)p(\btheta_2)$ in the combined data set $\bD$ and   $p_\ell(\btheta_1,\btheta_2)=p_\ell(\btheta_1)p_\ell(\btheta_2)$ in center $\ell$, for all $\ell$ (the ``$\ell$'' in the subscript refers to the center). We choose the prior parameter distributions for $\btheta_1$ and $\btheta_2$ to be Gaussian with mean zero and inverse covariance matrices $\bLambda_1$ and $\bLambda_2$, respectively, in the combined data set, and $\bLambda_{1,\ell}$ and $\bLambda_{2,\ell}$ in center $\ell$, $\ell=1,\ldots,L$.  For parameters that are positive by definition, like the variance of the error term in the linear regression model, a mean zero Gaussian prior is assumed for a transformation (e.g., the logarithm) of the parameter. 

The maximum a posteriori (MAP) estimate of $\btheta =(\btheta_1,\btheta_2)$ maximizes the a posteriori density of the data with respect to $\btheta$, by definition. For the combined data set $\bD$, this estimate is denoted as $\widehat{\btheta} = (\widehat{\btheta}_1,\widehat{\btheta}_2)$ and, for the local data set $\bD_\ell$ the notation $\widehat{\btheta}_\ell=(\widehat{\btheta}_{1,\ell},\widehat{\btheta}_{2,\ell})$ is used. If the prior density is chosen to be non-informative (large prior variances), the MAP estimates will be close to the maximum likelihood estimates. The estimator $\widehat\btheta$ is fictive as the data set $\bD$ can not be created. In the following we derive expressions for $\widehat{\btheta}$ in terms of the MAP estimators based on the local data sets $\bD_\ell$. Once the estimates in the separate centers have been found, these expressions tell us how to combine them to obtain (an approximation of) $\widehat{\btheta}$. 

For the fictive combined data set $\bD$ the log posterior density can be written as
\begin{align}
\label{eq:logpost}
\log \big\{p(\btheta|\bD)\big\} &=\; \log \Bigg\{\frac{p(\bD|\btheta) p(\btheta)}{p(\bD)} \Bigg\} \nonumber \\[4pt] 
&=\; \log \big\{p(\theta)\big\} \;+\; \log \big\{p(\bD|\btheta)\big\} - \log \big\{p(\bD)\big\} \nonumber \\[2pt]
&=\; \log \big\{p(\theta)\big\} \;+\; \sum_{\ell=1}^L \sum_{i=1}^{n_\ell}\log \big\{p(y_{\ell i}, \bx_{\ell i}|\btheta)\big\} - \log \big\{p(\bD)\big\} \nonumber\\
&=\; \log \big\{p(\theta_1)\big\} \;+\; \log \big\{p(\theta_2)\big\} \;+\; \sum_{\ell=1}^L \sum_{i=1}^{n_\ell}\log \big\{p(y_{\ell i}| \bx_{\ell i},\btheta_1)\big\}  \;+\; \sum_{\ell=1}^L \sum_{i=1}^{n_\ell}\log \big\{p(\bx_{\ell i}|\btheta_2)\big\} \;-\; \log \big\{p(\bD)\big\} 
\end{align}
by Bayes' rule (first equality), independence between the observations (third equality), and, among others, independence between $\btheta_1$ and $\btheta_2$ (fourth equality). Similarly, the logarithm of the posterior density in center $\ell$ can be written as 
\begin{align}
\log \big\{p_\ell(\btheta|\bD_\ell)\big\} &=\; \log \big\{p_\ell(\theta_1)\big\} \;+\; \log \big\{p_\ell(\theta_2)\big\} \;+\; \sum_{i=1}^{n_\ell}\log \big\{p(y_{\ell i}| \bx_{\ell i},\btheta_1)\big\} \;+\; \sum_{i=1}^{n_\ell}\log \big\{p(\bx_{\ell i}|\btheta_2)\big\} \;-\; \log \big\{p_\ell(\bD_\ell)\big\}.
\label{eq:loppostl}
\end{align}
The log posterior densities $\log \{p(\btheta|\bD)\}$ and $\log \{p_\ell(\btheta|\bD_\ell)\}$ are decomposed into terms that depend on either $\btheta_1$ or on $\btheta_2$, but never on both. As a consequence, maximization with respect to $\btheta_1$ and $\btheta_2$ to obtain their MAP estimators can be performed independently.  { By reordering the terms in expression (\ref{eq:loppostl}), we find 
\begin{align*}
\sum_{i=1}^{n_\ell}\log \big\{p(y_{\ell i}| \bx_{\ell i},\btheta_1)\big\} \;+\; \sum_{i=1}^{n_\ell}\log \big\{p(\bx_{\ell i}|\btheta_2)\big\} 
 \;=\; \log \big\{p_\ell(\btheta|\bD_\ell)\big\} \;-\; \log \big\{p_\ell(\theta_1)\big\} \;-\; \log \big\{p_\ell(\theta_2)\big\}  \;+\; \log \big\{p_\ell(\bD_\ell)\big\}.   
\end{align*}
The right hand side of this expression can be inserted into expression (\ref{eq:logpost}). Then, the log posterior density for the full data set $\log \{p(\btheta|\bD)\}$ is written as a sum of the local log posterior densities in the centers and the log prior densities (more details are given in Appendix II.A). For deriving the BFI estimators of the parameters, the local log posterior densities are approximated by second order Taylor expansions around the local MAP estimates. Instead of maximizing the full log posterior density for the combined data, the quadratic approximation is maximized with respect to the parameters. The parameter value where the maximum is attained is defined as the BFI estimate. } For $\widehat{\bA}_{1,\ell}$ and $\widehat{\bA}_{2,\ell}$ the second derivatives of $-\log \{p_\ell(\btheta|\bD_\ell)\}$ with respect to $\btheta_1$ and $\btheta_2$ and evaluated in the local MAP estimators $\widehat{\btheta}_{1,\ell}$ and $\widehat{\btheta}_{2,\ell}$, in center $\ell$,  the BFI estimators equal
\begin{align}
\hspace*{-5mm} & \widehat{\btheta}_{1,\BFI} :=  \big(\widehat{\bA}_{1,\BFI}\big)^{-1}\sum_{\ell=1}^L \widehat{\bA}_{1,\ell}\widehat{\btheta}_{1,\ell},~~~~~~~~~
\widehat{\bA}_{1,\BFI} := \sum_{\ell=1}^L \widehat{\bA}_{1,\ell}+\bLambda_1-\sum_{\ell=1}^L \bLambda_{1,\ell},
\label{eq:recover_theta1}
\\
\hspace*{-5mm} & \widehat{\btheta}_{2,\BFI} := \big(\widehat{\bA}_{2,\BFI}\big)^{-1}\sum_{\ell=1}^L \widehat{\bA}_{2,\ell}\widehat{\btheta}_{2,\ell}, ~~~~~~~~~ \widehat{\bA}_{2,\BFI} := \sum_{\ell=1}^L \widehat{\bA}_{2,\ell}+\bLambda_2-\sum_{\ell=1}^L \bLambda_{2,\ell},
\label{eq:recover_theta2}
\end{align}
see Appendix II.A for the derivation. With these expressions we can compute approximations of $\widehat{\btheta}_1$ and $\widehat{\btheta}_2$ {\em a posteriori} from the inference results on the subsets and there is no need to do inference on the (fictive) combined data set $\bD$ to find the BFI estimates.  In the calculations of the BFI estimators, we assume independence between the parameters $\theta_1$ and $\theta_2$. This assumption was made for mathematical convenience, as the log posterior density splits into terms that are a function of $\theta_1$ or of $\theta_2$, but never of both, and as a consequence, separate expressions for $\widehat{\theta}_{1,\BFI}$ and $\widehat{\theta}_{2,\BFI}$ are found. This independence assumption is not essential. If the parameters are dependent, the calculations can be performed in a similar way and a single expression for the BFI estimator for $(\theta_1,\theta_2)$ is found.

In Appendix III.B we prove that under the assumption of no model misspecification (including homogeneity between the centers), the BFI estimators $\widehat{\btheta}_{1,\BFI}$ and $\widehat{\btheta}_{2,\BFI}$ are asymptotically Gaussian and efficient (minimum asymptotic variance). For $n_\ell, \ell=1,\ldots,L$ the local sample sizes and $n=n_1+\ldots +n_L$ the total sample size, it is proven that
\begin{align*}
\sqrt{n}\big(\widehat{\btheta}_{1,\BFI}-\btheta_1\big) \leadsto {\mathcal N}\Big({\bf 0}, \big(\sum_{\ell=1}^L w_\ell I_{1,\ell}\big)^{-1}\Big),~~~~~~~~~~~~\mbox{with} ~~~~~~~~~~~w_\ell = \lim_{n_1,\ldots,n_\ell \rightarrow \infty} \frac{n_\ell}{n},    
\end{align*}
and $I_{1,\ell}$ the Fisher information matrix in center $\ell$ (the notation `$\leadsto$' means convergence in distribution). The matrix $\sum_{\ell=1}^L w_\ell I_{1,\ell}$ equals the Fisher information matrix for estimating $\btheta_1$ in the combined data set (see Appendix III.A). The BFI estimator asymptotically follows the same distribution as the MAP and the Maximum Likelihood estimators on the combined data. Apparently, no information is lost as a consequence of the fact that the data sets cannot be shared. In the homogeneous setting $I_{1,\ell}=I_1, \ell=1,\ldots,L$, independent of $\ell$, and $\sum_{\ell=1}^L w_\ell I_{1,\ell}=I_1$.
Further, since $n^{-1} \widehat{\bA}_{1,\BFI}$ converges in probability to $\sum_{\ell=1}^L w_\ell I_{1,\ell}$ (see Appendix III.B), the asymptotic covariance matrix can be estimated by the inverse of $n^{-1} \widehat{\bA}_{1,\BFI}$. Similar results hold for the BFI estimator $\widehat{\btheta}_{2,\BFI}$. 

It follows that for a sufficiently large total sample size, the BFI estimators $\widehat{\btheta}_{1,\BFI}$ and $\widehat{\btheta}_{2,\BFI}$ are approximately Gaussian with mean $\btheta_1$ and $\btheta_2$ and with covariance matrices that can be estimated by $\widehat{\bA}_{1,\BFI}^{-1}$ and $\widehat{\bA}_{2,\BFI}^{-1}$. From this, credible intervals for $\btheta_1$ and $\btheta_2$ can be constructed. Let $\btheta_{1,(k)}$ be the $k^{th}$ element of $\btheta_1$. This parameter is estimated by $\widehat{\btheta}_{1,\BFI (k)}$, the $k^{th}$ element of $\widehat{\theta}_{1,\BFI}$ and its approximate $(1-2\alpha) 100\%$ credible interval equals
$\widehat{\btheta}_{1,\BFI (k)} \pm \xi_\alpha \; \big(\widehat{\bA}_{1,\BFI}^{-1}\big)_{k,k}^{1/2},$
for $\xi_\alpha$ the upper $\alpha$-quantile of the standard Gaussian distribution and $\big(\widehat{\bA}_{1,\BFI}^{-1}\big)_{k,k}^{1/2}$ equal to the square root of the $(k,k)^{th}$ element of the inverse of the estimator $\widehat{\bA}_{1,\BFI}$. Hypothesis testing is also straightforward by the asymptotic normality.

\section{Heterogeneity across centers}
\label{sec: hetero}
In the derivation of the estimators for the aggregated BFI model in (\ref{eq:recover_theta1}) and (\ref{eq:recover_theta2}), homogeneity of the populations across the different centers is assumed. This assumption means that the parameters $\btheta_1$ and $\btheta_2$ are the same in every center.  This assumption may not be true, and the BFI approach has to be adjusted to take this heterogeneity into account. This is the topic of the present section.

In order to explain different types of heterogeneity, a specific example is used throughout the paper. This example is also used in Section \ref{sec: performance} and Appendix I to illustrate the BFI methodology and to study its performance. Here we give only a brief description, a more extensive description is given in Subsection \ref{subsec: Data description}. The example data come from a hypothetical study on stress among nurses on different wards in different hospitals \cite{Hox}. The data were simulated from a linear mixed effects model. The outcome of interest is job-related stress. For every nurse, information on stress, age, experience (in years), gender, wardtype (general, special care), hospital, and hospital size (small, medium, large) is available.  
 
Heterogeneity in the populations across  multiple centers may occur if, for instance, some medical centers are located in large cities and others in more rural areas; e.g., nurses in city medical centers may be younger on average than those in centers located in more rural areas. It might also be that in some hospitals the stress level among nurses is significantly higher than in others due to factors that are not nurse specific, like the size of the hospital or management decisions within a hospital (which are not in the data). In this section the following types of heterogeneity are considered:
\begin{enumerate}
\item Heterogeneity of population characteristics in the centers, e.g., the age distributions of the nurses differ. Then, the values of the parameter $\btheta_2$ differ across centers. This is considered in Subsection \ref{subsec: population}.
\item Heterogeneity across centers in outcome mean. This may happen if the  mean stress-level of the nurses vary across the centers due to factors that have not been measured (e.g., type of management). This is considered in Subsection \ref{subsec: location specific outcome}. 
\item Heterogeneity across centers due to interaction effects; the effect of a covariate varies across the centers. For instance, it might be that the effect of the wardtype on the outcome differs across medical centers. This means that the regression coefficient for wardtype is center-specific. This situation is considered in Subsection \ref{subsec: interaction}. 
\item Heterogeneity across centers due to center-specific nuisance parameters; e.g., the variance of the error term in a linear regression model. See Subsection \ref{subsec: nuisance}.
\item Heterogeneity across centers due to clustering; e.g., clustering by the location of the hospitals. This situation is considered in Subsection \ref{subsec: clustering}.
\item Heterogeneity across centers due to center-specific covariates. An example of such a covariate is hospital size, which is the same for every nurse in a hospital, but may vary across hospitals. See Subsection \ref{subsec: location specific covariate} 
\end{enumerate}

These types of between-center heterogeneity are due to center-specific parameters (types 1 to 4), due to clustering (type 5) and due to missing covariates (type 6). There may be more forms of heterogeneity that can be taken into account with the BFI methodology. The aim of the BFI approach is to increase the sample size relative to the parameter dimension to overcome overfitting. By significantly increasing the number of parameters in the BFI model, to account for heterogeneity,  the very objective of the BFI approach would thereby be undermined.

\subsection{Heterogeneity of population characteristics}
\label{subsec: population}
Characteristics of the populations who visit the $L$ centers may differ, for instance because the centers are located in different countries or regions. In the example, the fractions of female nurses differ across the centers. 

The parameter $\btheta$ was decomposed in $\btheta_1$ and $\btheta_2$. The parameter $\btheta_2$ describes the distribution of the covariates $\bX$, whereas the parameter $\btheta_1$ describes the relationship between the covariates and the outcome (so the regression coefficients and the nuisance model parameters). Under the assumption that $\btheta_1$ and $\btheta_2$ are independent,  the local log posterior densities were decomposed into terms that depend on either $\btheta_1$ or $\btheta_2$, but never on both (see expression (\ref{eq:loppostl})). As a consequence, when calculating the MAP estimates of $\btheta_1$ and $\btheta_2$, separate functions have to be maximized. Therefore, even if we would take into account that the populations vary across the centers, the BFI estimators $\widehat{\btheta}_{1,\BFI}$ and $\widehat{\bA}_{1,\BFI}$ in (\ref{eq:recover_theta1}) would not be affected,  and $\widehat{\btheta}_{1,\BFI}$ is still asymptotically efficient for estimating $\btheta_1$. For $\widehat{\btheta}_{2,\BFI}$ in (\ref{eq:recover_theta2}) new estimators need to be derived that take this heterogeneity into account. Since $\btheta_2$ is usually seen as a nuisance parameter and is not of main interest, this derivation is not included in the present paper.

\subsection{Heterogeneity across outcome means}
\label{subsec: location specific outcome}
If the combined data would be available for analysis, a multi-level model that includes a random center effect for possible unmeasured heterogeneity across centers would be considered. As an alternative one could include a fixed effect for the different centers. In both cases, this means that every center has its own center-specific intercept. At a local level, so within a center, it is not possible to estimate a center-effect. When combining the MAP estimators from the different centers into a BFI estimator for the combined model, different intercepts across the centers can be allowed in the model. This is explained below and the mathematical derivation can be found in Appendix II.B.

Suppose a regression model is fitted in every center based on the local data only. The BFI strategy as explained before, combines the fitted models to a model with a single general intercept. In Appendix II.B the BFI calculations are given for combining the local models in the situation that one or multiple regression parameters may vary across the centers and center-specific parameters are adopted in the aggregated BFI model. By taking this ``varying regression parameter'' to be the intercept in the resulting combined BFI model, every center has its own estimated intercept (and there is no general intercept). To be more specific, an estimate of the following aggregated BFI generalized linear model is obtained for an individual in center $\ell$
\begin{align}
h\big(\E (Y_{\ell i}| \bX_{\ell i}= \bx_{\ell i},\etab,\bbeta,\bgamma)\big) \;=\; \sum_{j=1}^{L} \beta_j 1_{\{\ell=j\}} + \bgamma^t \bx_{\ell i} \;=\; \beta_\ell + \bgamma^t \bx_{\ell i},     
\label{BFI model}
\end{align}
where the indicator function $1_{\{\ell=j\}}$ equals 1 if $\ell=j$ and 0 if $\ell \neq j$.  The parameters $\beta_1,\ldots,\beta_L$ are the center-specific intercepts and $\bgamma$ is the vector of regression parameters. The vector of covariates $\bx_{\ell i}$ does not include a 1 for the intercept. So, the aggregated BFI model for a nurse from center $\ell$ has an intercept $\beta_\ell$, which is specific for that center. 
The model can be easily rewritten into a form with a general intercept and parameters for the effect relative to the reference center which is taken to be center 1: 
\begin{align*}
h\big(\E (Y_{\ell i}| \bX_{\ell i}=\bx_{\ell i},\etab,\bbeta,\bgamma)\big) \;=\; \beta_1 + \sum_{j=2}^{L} \beta_j^\star 1_{\{\ell=j\}} + \bgamma^t \bx_{\ell i} \;=\; \beta_1 + \beta_\ell^\star + \bgamma^t \bx_{\ell i}, 
\end{align*}
where $\beta_\ell^\star = \beta_\ell - \beta_1$, for $\ell=2,\ldots,L$, with $\beta_\ell$ as in model (\ref{BFI model}). So, by allowing different intercepts when combining the fitted local models, the BFI model accounts for a "center-effect".

\subsection{Heterogeneity due to center interaction effects}
\label{subsec: interaction}
Next suppose that the effect of a covariate (a regression parameter) may vary across the centers. For instance, suppose that the effect of wardtype on job related stress may differ across the centers. In the regression model for the combined data, an interaction between the covariate wardtype and the hospital would be included. To obtain these estimates with the BFI approach, the calculations from Appendix II.B can be followed again, but this time for a regression parameter instead of the intercept. That gives an aggregated BFI model of the form: 
\begin{align*}
h\big(\E (Y_{\ell i}| \bX_{\ell i}=\bx_{\ell i}, z_{\ell i},\etab,\bbeta,\bgamma)\big) = \gamma_0 + \sum_{j=1}^{L} \beta_j ~ z_{\ell i} 1_{\{\ell=j\}} + \bgamma^t \bx_{\ell i},    
\end{align*}
where $\bgamma_0$ is the intercept, $\beta_j$ the wardtype effect on stress in center $j$, $z_{\ell i}$ the indicator function that indicates whether nurse $i$ from hospital $\ell$ is from a special care ward (0 general, 1 special care), $\bgamma$ the remaining regression parameters and $\bx_{\ell i}$ the vector of covariates (so without wardtype). %The term $\beta_\ell x_{\ell i, k}$ can be written as $\beta_\ell x_{\ell i, 1} = \beta_1 x_{\ell i, k} + \beta_\ell^\star x_{\ell i, k}$, where  $\beta_\ell^\star=\beta_\ell - \beta_1$, where $\beta_1$ is the parameter for center 1 and $\beta_\ell^\star$ is the relative effect compared to the effect in center 1.
%The presence of heterogeneity across the centers due to interaction effects can be verified by constructing credible intervals, as was described in Subsection \ref{subsec: location specific outcome}. 

\subsection{Heterogeneity due to having distinct nuisance parameters}
\label{subsec: nuisance}
The nuisance parameter of the statistical model, for example the variance of the error term in a linear regression model, may differ between the medical centers. Here too, the calculations for the BFI estimator in Appendix II.B can be applied. This yields an estimated aggregated BFI model with a specific nuisance parameter for each center.

\subsection{Heterogeneity due to center-clustering}
\label{subsec: clustering}
Local centers can be clustered based on, for example, geospatial regions, type of center (e.g., academic/non-academic hospital) or its size (small/medium/large). If the data can be combined, clustering can be taken into account by including a categorical variable in the model that represents this clustering. Within a center, this is not possible, because all persons in the center are in the same cluster and thus have the same variable value (which would lead to collinearity with the intercept); the regression model must be fitted without the corresponding variable. In this local model, the estimated intercept includes the clustering effect. When combining the models with the BFI approach, we must take this clustering into account. New expressions for the BFI estimators have been derived (Appendix II.C). For $K$ giving the number of clusters, the resulting BFI model has categorical specific intercepts:
\begin{align*}
h\big(\E (Y_{\ell i}| \bX_{\ell i}=  \bx_{\ell i},z_{\ell},\etab,\bbeta,\bgamma)\big) = \sum_{k=1}^K \beta_k 1_{\{z_{\ell}=k\}} + \bgamma^t \bx_{\ell i},     
\end{align*}
with $\beta_k$ the intercept for the $k^{th}$ cluster, $z_{\ell}$ represents the cluster of  center $\ell$, and $1_{\{z_{\ell}=k\}}$ is an indicator function that equals 1 if $z_{\ell}=k$ and 0 if $z_{\ell}\neq k$. As before, this model can be easily reformulated to a model with an intercept and a reference group.

\subsection{Heterogeneity due to center-specific covariates}
\label{subsec: location specific covariate}
Covariates that are included in the local models are also included in the aggregated BFI model. If a variable does not vary within a center (e.g., the size of the medical staff or the percentage of female patients) it can not be included in the regression model for the center and is, therefore, not automatically included in the BFI model. The effect of such a variable is then hidden in the intercepts of the local models. In this subsection we explain how the BFI approach can be adjusted to estimate a (combined) BFI model that includes this center-specific covariate. Although the problem is the same for categorical and continuous variables, the statistical solutions are not. This has to do with the way the variable is included in the aggregated BFI model. If the variable is categorical, one or more binary dummy variables need to be included in the model to represent every category (minus 1). If the variable is included in the model as a continuous variable, only one variable needs to be included (under the assumption of linearity) that holds for all centers.

If the variable is categorical and every center has its own specific category, we are in the situation as described in Subsection \ref{subsec: location specific outcome}, where the aggregated model has a center-specific intercept. If the number of categories is lower than the number of centers and multiple centers are in the same category, we actually have to deal with clustering as described in Subsection \ref{subsec: clustering}.

If the center-specific variable is continuous, for example the number of patients that is yearly treated in the corresponding center or the percentage of female patients, we actually want to fit a BFI model (based on all data) of the form:
\begin{align}
h\big(\E (Y_{\ell i}| \bX_{\ell i}=  \bx_{\ell i},z_{\ell},\etab,\nu_0,\nu_1,\bgamma)\big) = \nu_0 + \nu_1 z_\ell + \gamma^t \bx_{\ell i},     
\label{BFI cat model}
\end{align}
where $\nu_0$ is the intercept, $z_\ell$ is the continuous center-specific variable, and $\nu_1$ its corresponding unknown regression coefficient. The question is how to estimate the model parameters, and especially $\nu_0$ and $\nu_1$. This is explained below.

First all local models without this variable are fitted as described before. Next, the models are combined with the BFI methodology under the assumption that all intercepts may be different (the calculations are given in Appendix II.B and is also explained in Subsection \ref{subsec: location specific outcome}). This yields an estimate of the model with a center-specific intercept:
\begin{align*}
h\big(\E (Y_{\ell i}| \bX_{\ell i}= \bx_{\ell i},\etab,\bbeta,\bgamma)\big) \;=\; \beta_\ell + \bgamma^t \bx_{\ell i},     
\end{align*}
for center $\ell$. The effect of the continuous variable is hidden in this intercept: $\beta_\ell = \nu_0 + \nu_1 z_\ell$. To estimate $\nu_0$ and $\nu_1$ based on the estimated intercepts $\widehat{\beta}_\ell, \ell=1,\ldots,L$ and $z_\ell, \ell=1,\ldots,L$, one could make a scatter plot of the points $(z_1,\widehat{\beta}_1), \ldots, (z_L,\widehat{\beta}_L)$. Next, after fitting the least squares line through the points, the parameter $\nu_0$ can be estimated by the intercept of the least square line and $\nu_1$ by its slope.

%If there are multiple variables that are location-specific like the hospital size and whether it is an academic or non-academic hospital (this variable is not in our data set), a similar strategy can be applied. Important is that the model for the combined data would be identifiable. If this is not the case, for instance if all academic hospitals are large and all non-academic hospitals are of small or medium size, the model for the combined data is not identifiable and can not be estimated by combining the locally estimated model with the BFI methodology.

\subsection{Asymptotic performance of the BFI estimator under heterogeneity}
\label{subsec: asymp}

%So far, we have described different types of heterogeneity. For every setting it is explained how to deal with the heterogeneity when estimating the aggregated BFI model. Expressions for the corresponding BFI estimators have been derived in Appendix II. The settings in which the BFI estimators have been derived are formulated in a more general way: (1)  homogeneity between centers, (2) heterogeneity between centers due to center-specific parameters (3) heterogeneity between centers due to clustering. 

For both the homogeneous and the heterogeneous settings, the asymptotic distributions of the BFI estimators are derived in Appendix III. In the homogeneous setting, it turns out that the BFI estimator is asymptotically zero-mean Gaussian with covariance matrix equal to the inverse of the Fisher information matrix; the BFI estimator is asymptotically efficient. This distribution is equal to the asymptotic distribution of the MAP and maximum likelihood estimators that would have been based on the combined data; hence asymptotically no information is lost if the data cannot be merged.  

In the heterogeneous setting with center-specific parameters, the parameters of interest can be split into those that are the same between the centers and that are center-specific. Expressions of the corresponding BFI estimators are given in  (\ref{eq:theta1ahat}) and (\ref{eq:theta1bhat}) in Appendix II.B. In Appendix III.C it is proven that both BFI estimators are asymptotically Gaussian with covariance matrices that equal those for the MAP estimators and MLEs that would have been based on the combined data. Also in the heterogenous setting the BFI estimators are asymptotically efficient. Again asymptotically no information is lost if the data sets cannot be combined. In Appendix III.C it is proved that the BFI estimator for the center-specific parameter is asymptotically more accurate than the MAP estimator based on the local data of the center only. This is because the BFI estimator uses information from all centers to estimate the parameters that are the same across centers, while the MAP estimator uses local data only. A more accurate estimate of the shared parameters leads to a more accurate estimate of the non-shared parameters. 

Expressions of the BFI estimators for the setting in which the centers can be clustered are given in Appendix II.C. These expressions are complicated. Therefore, the derivation of the asymptotic distribution is not given here, but can be derived in the same way as for the setting with center-specific parameters.

Since the BFI estimator of $\btheta_1$ is asymptotically Gaussian and the asymptotic covariance matrix can be estimated by the inverse of $\widehat{\bA}_{1,\BFI}$, credible intervals can be easily constructed, as explained for the homogeneous setting. Hypotheses can be tested using the Wald test.

\subsection{Methods for checking heterogeneity}
In this paper we extend the BFI methodology to account for heterogeneity between centers. Before combining the local estimates, we should verify whether this heterogeneity is actually present and whether it is necessary to account for it. 

Suppose we want to investigate whether it is necessary to take into account the heterogeneity of the intercepts. Then, first the MAP estimates of the local intercepts, say $\widehat{\beta}_{\ell}, \ell=1,\ldots,L$, should be compared. However, there will always be differences between the estimates. The question is whether the observed differences are due to randomness or whether the true values of the intercepts are sufficiently different to take this into account in the modelling. The latter can be verified by constructing credible intervals. In order to compare the parameter estimates between two centers, say centers $k$ and $\ell$, a credible interval for the difference of the two intercepts can be constructed. Such a calculation is based on the statistical independence of the estimators $\widehat{\beta}_{k}$ and $\widehat{\beta}_{\ell}$ (since the data from the different centers are assumed to be independent) and the fact that $\widehat{\beta}_{k}$ and $\widehat{\beta}_{\ell}$ are approximately Gaussian with mean $\beta_k$ and $\beta_\ell$ and standard deviations $\big(\widehat{\bA}_{1,k}^{-1}\big)_{1,1}^{1/2}$ and $\big(\widehat{\bA}_{1,\ell}^{-1}\big)_{1,1}^{1/2}$, respectively, (if the first element of the parameter vectors $\btheta_{1,k}$ and $\btheta_{1,\ell}$ correspond to the intercept). Then, the $(1-2\alpha) 100\%$ credible interval for the difference $\beta_k-\beta_\ell$ equals 
\begin{align*}
\widehat{\beta}_{k}-\widehat{\beta}_{\ell} \;\pm\; \xi_\alpha \; \sqrt{\Big(\widehat{\bA}_{1,k}^{-1}\big)_{1,1}+\big(\widehat{\bA}_{1,\ell}^{-1}\big)_{1,1}}, 
\end{align*}
for $\xi_\alpha$ equal to the upper $\alpha$-quantile of the standard Gaussian distribution. With the latter interval we can verify whether the parameters in the centers $k$ and $\ell$ are different. If the sample sizes in the centers are small, the credible intervals may be wide and it may be difficult to conclude on hetereogeneity. 

Similarly, the $(1-2\alpha) 100\%$ credible intervals for the difference between the true $\beta$-value in all centers except $\ell$ and the true parameter value in center $\ell$ equals:
\begin{align*}
\widehat{\beta}_{-\ell,\BFI}-\widehat{\beta}_{\ell} \; \pm \; \xi_\alpha \; \sqrt{\big(\widehat{\bA}_{1,\BFI,-\ell}^{-1}\big)_{1,1}+\big(\widehat{\bA}_{1,\ell}^{-1}\big)_{1,1}},   
\end{align*} 
where subscript $-\ell$ means that the BFI estimator was computed not including the estimator from center $\ell$. With this interval we can verify whether the intercept in center $\ell$ differs from the intercepts in the other centers assuming that these intercepts equal.

In the same way, one can check whether it is necessary to take into account any of the other types of heterogeneity.

\section{Performance of BFI methodology}
\label{sec: performance}
The BFI methodology for GLMs was introduced in Jonker et al\cite{Jonker} and extended to survival models for homogeneous populations in Pazira et al\cite{Pazira}. Simulation studies in those papers show good performance of the methodology in the homogeneous setting. In this paper we focus on different types of heterogeneity. The results of simulation studies (Subsection \ref{subsec: sim study}) and data analyses (Subsection \ref{subsec: data analysis}) are described below.   

%The nurse-data mentioned earlier are used for the analysis and the simulation study. In Subsection \ref{subsec: Data description} we give a full description of the data. In Subsection \ref{subsec: Data analysis} we analyze the data with the BFI approach and compare the BFI estimates to those obtained with the combined data. Next, in Subsection \ref{subsec: prediction} we study the performance of the BFI methodology for predicting the outcome in case of homogeneity and heterogeneity across populations. 

%In this section, we assume the variance-covariance matrices of the Gaussian prior distributions $\Lambda$ and $\Lambda_\ell, \ell=1,\ldots,L$ to be diagonal matrices with diagonal elements equal to $\lambda$. That means that we assume that the parameters are independent and their variances are equal. We do the analyses and simulation studies for different values $\lambda$ to study the effect of the chosen value. 

\subsection{Simulation Studies}
\label{subsec: sim study}

\subsubsection{One-shot estimators for comparison}
As explained in the introduction, we are only interested in one-shot estimators, i.e., estimators that can be calculated after a single communication with the centers, like the BFI estimator. To enable performance comparison for the BFI estimator, we consider two one-shot estimators. The most interesting one is the weighted average estimator (WAV) which is defined as the weighted average of the local MAP estimators with the weights equal to $n_\ell/n$ (where $n=\sum_{\ell=1}^{L} n_\ell$); estimates based on larger data-sets are given larger weights. The weighted average estimator for $\btheta$ is defined as: 
\begin{align*}
%\label{form:betanaive}
\widehat{\btheta}_{\WAV}  &=  \sum_{\ell=1}^{L} \frac{n_\ell}{n} ~\widehat{\btheta}_{\ell}.
\end{align*}
In case of clustering, the WAV estimator for the parameter that is specific for a particular cluster is defined as the weighted average of the local MAP estimators of the centers in that cluster. If a parameter may vary between all centers, the corresponding WAV estimator is defined as the MAP estimator in the local center. The second one-shot estimator for $\btheta$ is the single center estimator $\widehat{\btheta}_{\single}$, defined as the MAP estimator in the center with the largest local sample size. The single center estimator cannot be defined in case of center or cluster specific parameters. 

In Appendix III the asymptotic distributions of the weighted average and the single center estimators are derived. As expected, the asymptotic variance of the single center estimator is larger than the one of the BFI estimator,  because it is based on fewer data points. In the homogeneous setting, the WAV estimator turns out to be asymptotically efficient (minimum variance) and  it follows asymptotically the same distribution as the BFI estimator. In the heterogeneous setting, the WAV estimator of the parameter that differs between the centers is defined as the (corresponding) single center estimator. As explained in Subsection \ref{subsec: asymp}, the BFI estimator has a smaller asymptotic variance than this estimator. In this section the finite sample behaviour of the estimators are  compared by means of simulation studies.

\subsubsection{Performance measures for finite samples}
%{\bf Asymptotic distribution}\\
%{\bf In Appendix III, the asymptotic behaviour of the three estimators is computed and compared. Their convergence rates equal, and they are all asymptotically Gaussian, but the asymptotic covariance matrices differ. It is shown that the BFI estimator is asymptotically efficient and more accurate than the weighted average and the single center estimator. Only in the homogeneous setting in which the asymptotic distributions (actually the asymptotic covariance matrices) of the local MAP estimators equal, the weighted average estimator is asymptotically efficient as well. }

\noindent
Since the BFI methodology tries to reconstruct from local inferences what would have been obtained if the data sets had been merged, the BFI estimators by definition cannot do better than the MAP estimators based on the combined data. Therefore, the parameter estimates and outcome predictions obtained by the BFI approach are compared to those found after combining the data. For completeness, we also compare the estimates with the true parameter values. 

In the next subsection, the simulation procedure is explained. In brief, $B$ times data sets are simulated from a chosen model, for every center separately. In every cycle the parameters are estimated with the three one-shot estimators, and also by the MAP estimator based on the combined data. Performance is measured with the $\MSER_{\theta_k, \BFI}$ defined as 
\begin{align*}
\MSER_{\theta_k, \BFI}  = \frac{1}{B} \sum_{b=1}^{B}  \big(\widehat{\btheta}_{\BFI, k}^{(b)} - \widehat{\btheta}_{\com, k}^{(b)} \big)^2,
\end{align*}
where $\widehat{\btheta}_{\BFI, k}^{(b)}$ is the BFI estimated value of the $k^{\text{th}}$ coordinate of $\btheta$ in the $b^{\text{th}}$ iteration, and $\widehat{\btheta}_{\com, k}^{(b)}$ the estimate using all data. The MSE's for the other estimators are defined similarly:
\begin{align*}
\MSER_{\theta_k, \WAV}  = \frac{1}{B} \sum_{b=1}^{B}  \big(\widehat{\theta}_{\WAV, k}^{(b)} - \widehat{\theta}_{\com, k}^{(b)} \big)^2, ~~~~~~~~~~~ \MSER_{\theta_k, \single}  = \frac{1}{B} \sum_{b=1}^{B}  \big(\widehat{\btheta}_{\single, k}^{(b)} - \widehat{\btheta}_{\com, k}^{(b)} \big)^2.
\end{align*}
If the MSE is small, the estimates based on the local inference results are similar to the estimates based on the combined data, and thus only little information is lost. For the BFI estimator we also computed the MSE compared to the true parameter value; denoted as $\MSET_{\theta_k, \BFI}$ (where the $T$ stands for ``true value'').

\subsubsection{Simulation settings and results}
We assume that there are four centers with data of $n_1, n_2, n_3$ and $n_4$ individuals. For each individual, data of three independent covariates are simulated: two from a Gaussian distribution and one from a binomial distribution. The outcome variables given the covariates are assumed to be independent and are simulated from a logistic regression model. We consider the following situations: 1) the populations are homogeneous, 2) the distributions of the covariates differ across the centers, 3) the intercepts (prevalence) differ across the centers, and 4) centers are clustered. For the sample sizes we consider two settings: small sample sizes ($n_1=n_2=50$, $n_3=n_4=100$) and large sample sizes ($n_1=n_2=100$, $n_3=n_4=200$) and we set the covariance matrices of the Gaussian prior equal to diagonal matrices with $\lambda=0.001$ or $\lambda=0.01$ (or a mix) on the diagonal. This corresponds to variances that equal $1000$ and $100$ respectively; the prior distributions are almost non-informative. The first covariate is sampled from a Gaussian distribution with mean zero and standard deviation equal to $1$. The second covariate is Gaussian as well, but with mean $2$ and  standard deviation $5$. The third covariate comes from a binomial distribution with probability $0.25$. In the setting with heterogeneous populations, different covariate distributions have been used across the centers (see the caption of Table \ref{Tab: Table2}). In all cases the regression parameters equal  $1.0$ for the intercept and $2.0, -1.0$, and $0.5$ for the three covariates. 

For every setting, we simulate $B=1000$ data sets, compute the BFI, weighted average and single center estimates (the latter one only if relevant), and compute the MSEs. The simulation results in the four different settings  are given in Table \ref{Tab: Table1} (homogeneity between centers), Table \ref{Tab: Table2} (different covariate distributions), Table \ref{Tab: Table3} (different intercepts between centers) and Table \ref{Tab: Table4} (clustering).

\begin{table}%[t!]
\centering
\scalebox{0.65}{
\begin{tabular}{|c|c|r|r|r|r|r|r|r|r|r|r|r|r|r|r|r|r|}
		\hline
		& & \multicolumn{4}{c|}{$10^2 \times \MSER_{\bbeta, \BFI}$} & \multicolumn{4}{c|}{$10^2 \times \MSER_{\bbeta, \WAV}$} & \multicolumn{4}{c|}{$10^2 \times \MSER_{\bbeta, \single}$} & \multicolumn{4}{c|}{$10^2 \times \MSET_{\bbeta, \BFI}$}  \\ %\cmidrule{4-7}\cmidrule{9-12}\cmidrule{14-17}\cmidrule{19-22}
		$(n_1,n_2,n_3,n_4)$ & $(\lambda_{12},\lambda_{34})$  & $\beta_1$ & {$\beta_2$} & $\beta_3$ & {$\beta_4$} & $\beta_1$ & {$\beta_2$} & $\beta_3$ & {$\beta_4$} & $\beta_1$ & {$\beta_2$} & $\beta_3$ & {$\beta_4$} & $\beta_1$ & {$\beta_2$} & $\beta_3$ & {$\beta_4$}  \\
\hline
(25,25,50,50) & (0.001, 0.001) &  14.42 & 41.14 & 9.81 & 24.52 &   496.6 & 1322 & 387.6 & 325.4 &   997.9 & 2495 & 636.5 & 1069 &   21.36 & 35.42 & 6.81 & 59.52\\

%\cmidrule{2-22}
& (0.01, 0.01) &   10.83 & 32.6 & 8.11 & 14.92 &  79.93 & 198.8 & 65.92 & 64.8 &   229.5 & 432.2 & 103.4 & 387.2 &   21.03 & 36.28 & 6.62 & 64.46\\

%\cmidrule{2-22}
& (0.01, 0.001) &  12.76 & 36.91 & 9.36 & 17.37 &   271.7 & 760.1 & 219.9 & 224.8 &   1234 & 2485 & 686.2 & 1136 &   20.60 & 34.99 & 6.66 & 62.83\\

\hline 
(50,50,100,100) & (0.001, 0.001) & 3.54 & 11.47 & 2.74 & 3.73   & 74.45 & 225.8 & 63.5 & 77.21 &   62.93 & 279.17 & 51.37 & 196.6 &   8.38 & 14.58 & 2.72 & 24.62\\
%\cmidrule{2-22}
& (0.01, 0.01) & 3.51 & 9.91 & 2.55 & 2.73 &   18.50 & 45.73 & 12.43 & 18.67 &   44.10 & 81.51 & 15.89 & 94.96 &   8.56 & 14.02 & 2.62 & 24.50\\
%  \cmidrule{2-22}
& (0.01, 0.001) &  3.60 & 10.89 & 2.74 & 3.33 &   19.72 & 57.89 & 15.27 & 20.66 &   50.25 & 119.2 & 25.19 & 105.9 &   9.14 & 14.22 & 2.54 & 26.64\\
\hline
(100,100,200,200) & (0.001, 0.001) &  0.77 & 2.49 & 0.61 & 0.54 &   3.25 & 10.49 & 2.25 & 2.87 &   10.80 & 16.10 & 3.05 & 31.73 &   4.39 & 6.32 & 1.11 & 11.72\\
%\cmidrule{2-22}
& (0.01, 0.01) & 0.80 & 2.64 & 0.64 & 0.60 &   2.21 & 6.94 & 1.64 & 2.76 &   10.92 & 19.25 & 3.08 & 34.88 &   4.21 & 6.30 & 1.07 & 11.92\\
%\cmidrule{2-22}
& (0.01, 0.001) &  0.82 & 2.72 & 0.66 & 0.60 &   2.12 & 6.60 & 1.52 & 2.22 &   12.78 & 21.11 & 3.87 & 34.10 &   4.43 & 6.25 & 1.08 & 13.02\\
\hline
(200,200,400,400) & (0.001, 0.001) &   0.17 & 0.60 & 0.14 & 0.12 &   0.27 & 0.88 & 0.21 & 0.39 &   4.72 & 7.16 & 1.08 & 14.48 &   2.16 & 3.02 & 0.49 & 6.21\\

%\cmidrule{2-22}
& (0.01, 0.01) &  0.18 & 0.59 & 0.14 & 0.11 &   0.29 & 0.87 & 0.21 & 0.35 &   5.16 & 7.26 & 1.21 & 14.96   & 2.04 & 2.93 & 0.48 & 6.91\\

%\cmidrule{2-22}
& (0.01, 0.001) &  0.20 & 0.65 & 0.16 & 0.14 &  0.30 & 0.94 & 0.23 & 0.39 &   5.30 & 7.31 & 1.22 & 16.31 &   2.00 & 2.87 & 0.50 & 6.53\\

\hline 
\end{tabular}}
\caption{ { Homogeneous setting. The MSEs for the BFI, weighted average and the single center estimators, and MSET for the BFI estimator. The prior inverse covariance matrices are diagonal with the diagonal element equal to $\lambda_{12}$ in centers 1 and 2, and the value $\lambda_{34}$ in centers 3 and 4. The single center estimates are based on data from center 4 only. }}
\label{Tab: Table1}
\end{table}

\begin{table}%[t!]
\centering
\scalebox{0.65}{
\begin{tabular}{|c|c|r|r|r|r|r|r|r|r|r|r|r|r|r|r|r|r|}
		\hline
		& & \multicolumn{4}{c|}{$10^2 \times \MSER_{\bbeta, \BFI}$} & \multicolumn{4}{c|}{$10^2 \times \MSER_{\bbeta, \WAV}$} & \multicolumn{4}{c|}{$10^2 \times \MSER_{\bbeta, \single}$} & \multicolumn{4}{c|}{$10^2 \times \MSET_{\bbeta, \BFI}$}  \\ %\cmidrule{4-7}\cmidrule{9-12}\cmidrule{14-17}\cmidrule{19-22}
		$(n_1,n_2,n_3,n_4)$ & $(\lambda_{12},\lambda_{34})$  & $\beta_1$ & {$\beta_2$} & $\beta_3$ & {$\beta_4$} & $\beta_1$ & {$\beta_2$} & $\beta_3$ & {$\beta_4$} & $\beta_1$ & {$\beta_2$} & $\beta_3$ & {$\beta_4$} & $\beta_1$ & {$\beta_2$} & $\beta_3$ & {$\beta_4$}  \\
\hline
(50,50,100,100) & (0.001, 0.001) &  3.22 & 8.96 & 2.28 & 3.27 &   133.8 & 169.9 & 38.34 & 100.3 &   1093 & 1245 & 272.2 & 879.9   & 14.33 & 9.93 & 2.76 & 19.63\\
%\cmidrule{2-22}
& (0.01, 0.01) &  3.31 & 8.55 & 2.27 & 2.89 &   64.24 & 46.38 & 10.43 & 35.90 &  534.5 & 242.9 & 53.92 & 340.5 &   14.71 & 9.35 & 2.75 & 20.03\\
%\cmidrule{2-22}
& (0.01, 0.001) &  3.01 & 8.20 & 2.09 & 2.65 &   107.7 & 128.2 & 27.23 & 73.06 &   955.5 & 923.5 & 192.5 & 668.4 &   14.62 & 9.84 & 2.59 & 18.75\\
\hline
(100,100,200,200) & (0.001, 0.001) &  0.64 & 1.82 & 0.49 & 0.53 &   14.99 & 6.15 & 1.10 & 9.42 &   125.5 & 38.34 & 6.43 & 115.4 &   8.30 & 4.09 & 1.21 & 10.13\\
%\cmidrule{2-22}
& (0.01, 0.01) &  0.66 & 1.84 & 0.50 & 0.51 &   16.26 & 7.50 & 1.37 & 8.21 &   148.7 & 49.35 & 9.34 & 100.80 &   7.48 & 4.03 & 1.16 & 9.74\\
%\cmidrule{2-22}
& (0.01, 0.001) &  0.60 & 1.87 & 0.49 & 0.63 &   17.03 & 7.15 & 1.35 & 11.62 &   149.1 & 50.89 & 9.37 & 134.70 &   7.86 & 4.20 & 1.28 & 9.81\\
\hline  
\end{tabular}}
\caption{ { Heterogeneous setting. The MSEs for the BFI, weighted average and the single center estimators, and MSET for the BFI estimator. The distributions of the covariates differ across the centers. The first covariate is Gaussian with mean $0, 1, 2$, and $3$ in the four centers, and standard deviation $1$. The second covariate is Gaussian as well with mean $2$, but now the standard deviation varies: $1, 2, 3$, and $4$ in the four centers. The third covariate comes from a binomial distribution with probability $0.35, 0.30, 0.25$ and $0.20$ in the four centers. In all cases the prior inverse covariance matrix equals diagonal matrices with the diagonal element equal to $\lambda_{12}$ in the centers 1 and 2, and the value $\lambda_{34}$ in the centers 3 and 4. The single center estimates are based on data from center 4 only.}}
\label{Tab: Table2}
\end{table}

\begin{table}%[t!]
\centering
\scalebox{.6}{
\begin{tabular}{|c|r|r|r|r|r|r|r|r|r|r|r|r|r|r|r|r|r|r|r|r|r|}
		\hline
		 & \multicolumn{7}{c|}{$10^2 \times \MSER_{\bbeta, \BFI}$} & \multicolumn{7}{c|}{$10^2 \times \MSER_{\bbeta, \WAV}$} & \multicolumn{7}{c|}{$10^2 \times \MSET_{\bbeta, \BFI}$}    \\ %\cmidrule{2-8}\cmidrule{10-16}\cmidrule{18-24}
		 $(\lambda_{12},\lambda_{34})$  & $\beta_1$ & \multicolumn{1}{c}{$\beta_2$} & $\beta_3$ & \multicolumn{1}{c}{$\beta_4$} & $\beta_5$ & $\beta_6$ & $\beta_7$ & $\beta_1$ & \multicolumn{1}{c}{$\beta_2$} & $\beta_3$ & \multicolumn{1}{c}{$\beta_4$} & $\beta_5$ & $\beta_6$ & $\beta_7$ & $\beta_1$ & \multicolumn{1}{c}{$\beta_2$} & $\beta_3$ & \multicolumn{1}{c}{$\beta_4$} & $\beta_5$ & $\beta_6$ & $\beta_7$  \\
 \hline
 (0.001, 0.001) & 7.45 & 10.29 & 2.41 & 4.21 & 14.33 & 3.60 & 5.90 &   703.17 & 1193.88 & 24.94 & 163.43 & 352.73 & 94.05 & 103.72 &   42.35 & 44.94 & 19.27 & 22.94 & 12.47 & 3.04 & 32.63\\
%\cmidrule{1-24}
 (0.01, 0.01) & 4.00 & 6.66 & 1.96 & 3.39 & 12.41 & 3.08 & 4.10 &   120.80 & 266.14 & 21.43 & 46.34 & 73.79 & 19.59 & 25.15 &   41.36 & 42.99 & 20.09 & 20.96 & 11.7 & 2.82 & 34.23\\
%\cmidrule{1-24}
 (0.01, 0.001) & 4.77 & 7.76 & 2.49 & 4.63 & 14.04 & 3.48 & 5.29 &   130.48 & 241.2 & 51.51 & 230.56 & 186.45 & 50.05 & 71.33 &   36.99 & 40.48 & 19.13 & 23.19 & 11.47 & 2.71 & 33.52\\
\hline
  (0.001, 0.001) & 0.72 & 1.36 & 0.44 & 0.76 & 3.19 & 0.80 & 0.82 &   37.03 & 80.79 & 4.82 & 8.34 & 19.85 & 5.08 & 7.32 &   18.80 & 19.56 & 9.64 & 9.92 & 5.24 & 1.27 & 15.63\\
%\cmidrule{1-24}
 (0.01, 0.01)& 0.44 & 1.20 & 0.36 & 0.75 & 3.02 & 0.75 & 0.96 &   19.57 & 46.91 & 4.93 & 7.26 & 8.88 & 2.26 & 4.27 &   17.85 & 19.10 & 9.79 & 10.87 & 5.49 & 1.30 & 16.05\\
%\cmidrule{1-24}
 (0.01, 0.001)&0.45 & 1.30 & 0.41 & 0.79 & 3.14 & 0.78 & 0.91 &   25.72 & 47.00 & 5.71 & 8.38 & 11.27 & 2.81 & 4.72 &   17.35 & 19.37 & 9.51 & 11.59 & 5.26 & 1.19 & 15.77\\
\hline  
\end{tabular}}
\caption{ { Heterogeneous setting. The MSEs for the BFI and weighted average estimators, and MSET for the BFI estimator. The intercepts differ across the centers. The parameters  $\beta_1, \ldots, \beta_4$ are the center-specific intercepts for the four centers (with true values $0.0, 1.0, 0.5$ and $-1.0$). The parameters $\beta_5,\ldots,\beta_7$ are the regression coefficients for the three covariates. For the upper three lines in the table, the local sample sizes equal $(n_1, n_2, n_3, n_4) = (50, 50, 100, 100)$, and for the lower three lines they equal $(n_1, n_2, n_3, n_4) = (100, 100, 200, 200)$. The MSE for the single center estimator has been left out, because this estimator can estimate one intercept only.}}
\label{Tab: Table3}
\end{table}

\begin{table}%[t!]
\centering
\scalebox{.7}{
\begin{tabular}{|c|c|r|r|r|r|r|r|r|r|r|r|r|r|r|r|r|r|}
		\hline
		& & \multicolumn{5}{c|}{$10^2 \times \MSER_{\bbeta, \BFI}$} & \multicolumn{5}{c|}{$10^2 \times \MSER_{\bbeta, \WAV}$}  & \multicolumn{5}{c|}{$10^2 \times \MSET_{\bbeta, \BFI}$}  \\ %\cmidrule{4-8}\cmidrule{10-14}\cmidrule{16-20}
		$(n_1,n_2,n_3,n_4)$ & $(\lambda_{12},\lambda_{34})$  & $\beta_1$ & \multicolumn{1}{c}{$\beta_2$} & $\beta_3$ & \multicolumn{1}{c}{$\beta_4$} & $\beta_5$ & $\beta_1$ & \multicolumn{1}{c}{$\beta_2$} & $\beta_3$ & \multicolumn{1}{c}{$\beta_4$} & $\beta_5$ & $\beta_1$ & \multicolumn{1}{c}{$\beta_2$} & $\beta_3$ & \multicolumn{1}{c}{$\beta_4$} & $\beta_5$  \\
\hline
(50,50,100,100) & (0.001, 0.001) &  8.17 & 14.82 & 13.69 & 3.38 & 5.45 &   814.44 & 151.48 & 333.89 & 86.33 & 114.08 &   24.23 & 20.82 & 11.71 & 2.97 & 30.59\\
%\cmidrule{2-20}
& (0.01, 0.01) &  5.24 & 12.91 & 12.21 & 3.03 & 4.24 &   126.42 & 44.30 & 62.15 & 16.21 & 26.07 &   23.92 & 21.60 & 12.55 & 2.94 & 32.55\\
%\cmidrule{2-20}
& (0.01, 0.001) &  5.55 & 13.08 & 12.35 & 3.14 & 4.35 &   136.91 & 105.72 & 92.36 & 24.71 & 31.79 &   23.71 & 22.37 & 12.45 & 3.06 & 34.42\\
\hline
(100,100,200,200) & (0.001, 0.001) &  1.22 & 3.34 & 3.19 & 0.78 & 0.90 &   44.17 & 6.49 & 15.55 & 3.79 & 7.41 &   9.59 & 9.54 & 4.89 & 1.11 & 14.02\\
%\cmidrule{2-20}
& (0.01, 0.01) & 1.19 & 3.07 & 2.97 & 0.73 & 0.72 &   24.54 & 6.49 & 8.54 & 2.12 & 3.79 &   10.06 & 9.56 & 5.16 & 1.25 & 14.4\\
%\cmidrule{2-20}
& (0.01, 0.001) &  1.30 & 3.38 & 3.21 & 0.79 & 0.75 &   26.25 & 7.23 & 9.38 & 2.24 & 3.65 &   10.54 & 10.28 & 5.45 & 1.31 & 15.02\\
\hline  
\end{tabular}}
\caption{  { Heterogeneous setting. The MSEs for the BFI and weighted average estimators, and MSET for the BFI estimator. The centers 1 and 2 and the centers 3 and 4 are clustered. The parameter $\beta_1$ and $\beta_2$ are the cluster specific intercepts (true values $1.0$ and $2.0$). The parameters $\beta_3, \beta_4$ and $\beta_5$ are the regression parameters of the three covariates. The MSE for the single center estimator has been left out, because the intercept for a single cluster can be estimated.}}
\label{Tab: Table4}
\end{table}

From the results in the tables it can be seen that for all estimators the MSEs  decrease for increasing sample size. For the BFI estimator the decrease is stronger for the MSEs than for the MSETs. A decrease is as expected as a larger sample size yields more accurate estimates.

Further, the results show that the MSEs for the BFI estimates are smaller than those for the weighted average and the single-center estimates. This also holds MSET (the $\MSET_{\beta, \WAV}$ and $\MSET_{\beta, \single}$ are left out from the tables, due to a lack of space.) The relative differences between $\MSER_{\beta, \BFI}$ and $\MSER_{\beta, \WAV}$ decrease with increasing sample size. This was expected, as (in the homogeneous setting) the asymptotic distributions of the BFI and the WAV estimators are identical. For finite sample sizes the differences in MSE are still considerable. 

In all settings the $\MSER_{\beta, \BFI}$ is smaller than the $\MSET_{\beta, \BFI}$. This is as expected, as the randomness in the observations is reflected in the estimate based on the combined data, but not in the actual parameter values. An important observation is that within every setting and for all combinations of sample sizes, the $\MSET_{\beta, \BFI}$ is rather stable for the different combinations of $\lambda$-values. Since the actual parameter values are independent of the choice of $\lambda$, we can conclude that the BFI estimates are not very sensitive to the values of $\lambda$ (considered here). However, the MSE for the three estimators decreases (slightly) for increasing $\lambda$ (a smaller prior variance), especially when the sample size is small. For larger values of $\lambda$, the MAP estimates have shrunk further to zero, leading to smaller MSEs. The latter does not imply that the estimates are more similar to the actual values. 

When comparing the MSEs of the different regression parameters (within the same setting and set of sample sizes), it is clear that some regression parameters can be estimated more accurately than others. For example, comparing the MSEs for the regression coefficients of the first and the second covariate (i.e., for $\beta_2$ and $\beta_3$ in Table \ref{Tab: Table2}) it can be seen that the MSEs for the coefficients for the second covariate are smaller, probably because the variation in this covariate is larger than in the first one. This applies to all estimators.

When comparing the values of $\MSET_{\beta, \BFI}$ in tables \ref{Tab: Table1} and \ref{Tab: Table2}, we see that the estimates of the regression parameters (except the intercept) are more accurate in the heterogeneous setting with different covariate distributions across the centers. Again, this may be due to the increased variation in the covariate values. 

Also in settings with center-specific intercepts (different prevalence in the centers) and clustering, the BFI estimators clearly perform better than the weighted average estimators, this is especially true for the center-specific and the cluster specific intercepts. Within the BFI methodology, these estimates use information from all centers for estimation. This is not the case for the weighted average estimator.

\subsection{Data analysis}
\label{subsec: data analysis}
\subsubsection{Description of the data}
\label{subsec: Data description}
The data come from a hypothetical study on stress among nurses in hospitals\cite{Hox}. The data set consists of simulated data of 1000 nurses working on different wards in 25 hospitals.\footnote{The data are available in the software package {\tt BFI} in R.} The outcome of interest is job-related stress among nurses. Additionally, for every nurse the following variables are available: age (years), experience (years), gender (0 = male, 1 = female), the type of ward in which the nurse works (0 = general care, 1 = special care), hospital (1, 2, \ldots, 25), and hospital size (small, medium, large). The number of nurses per hospital runs from 36 to 52.  For the covariates, the average age in the hospitals runs from 39.2 to 46.3 years, the fraction of female nurses from 0.61 to 0.85, the number of years of experience from 14.9 to 18.5, and the fraction of nurses on a special care ward runs from 0.48 to 0.51. For some of these variables there is hardly any variation across the hospitals, whereas for other variables the variation is much larger, like the fraction of female nurses. So, there seems to be some heterogeneity of the population characteristics across the centers (see Subsection \ref{subsec: population}). Further, there are 9 small hospitals, 12 medium sized hospitals and 4 large hospitals. The average stress level in hospitals increases with the size of the hospital; there is heterogeneity due to a hospital size clustering effect (see Subsection \ref{subsec: clustering}). Also the average stress level varies across the centers; between 3.6 and 5.8. Part of this heterogeneity may be due to non-measured hospital effects like location and patient population (see subsections \ref{subsec: location specific outcome} and \ref{subsec: location specific covariate}). In every hospital we fitted a linear regression model and  estimated the variance of the error term. The estimated variances vary from 0.17 to 1.16. It seems that there may be heterogeneity in this variance parameter (see Subsection \ref{subsec: nuisance}). In Subsection \ref{subsec: Data analysis} we estimate linear regression models with the BFI methodology, adjusted for these types of heterogeneity.  

For better comparison and interpretation of the estimates of the regression parameters, we standardized the continuous variables age, experience and stress: from each observed value we subtracted the full sample mean and divided the result by its full sample standard deviation. This is not required for the BFI method. However, note that this can be easily done without combining all data, since the full sample mean and standard deviation can be easily reconstructed from the local sample means and local standard deviations (and thus only these values need to be shared with the central server).

\subsubsection{Model estimation under heterogeneity}
\label{subsec: Data analysis}
In this subsection we analyse the data from the 25 centers with the BFI methodology and we compare the estimated aggregated BFI model to the model that would have been found if the data had been combined before fitting the model. As described in the previous subsection we have different types of heterogeneity. We start with a relatively simple linear regression model and combine the local MAP estimates with the BFI methodology under the assumption of homogeneity across the centers. In a second analysis we also include a clustering effect for the variable hospital size, in the third analysis we allow a center-specific intercept, and in the last step we also allow for difference variances of the error term. In Appendix I it is explained how these analyses can be performed in R with our R-package {\tt BFI}.

In the first analysis we only include nurse-specific variables: age, gender, experience (exp), and wardtype. We fit a linear regression model of the form:
\begin{align*}
\mbox{stress}_{\ell i} ~=~ \beta_0 ~+~ \beta_1~ \mbox{age}_{\ell i} ~+~  \beta_2~ \mbox{gender}_{\ell i} ~+~ \beta_3~ \mbox{exp}_{\ell i}  ~+~ \beta_4~ \mbox{wardtype}_{\ell i} ~+~ \varepsilon_{
\ell i},
\end{align*}
where the subscript ``$\ell i$'' refers to the $i^{th}$ person in center $\ell$. The last term, $\varepsilon_{\ell i}$, is the measurement error in the outcome variable, which is assumed to be Gaussian with mean zero and variance $\sigma^2$. In the analyses based on the combined data and in the centers we take Gaussian priors with a diagonal inverse covariance matrix $\Lambda$ with either $\lambda=0.001$ or $\lambda=0.1$ on the diagonal. For these values of $\lambda$ the corresponding variances of the parameter priors are equal to $1000$ and $10$, respectively. For a prior variance equal to $1000$, the MAP estimates are close to the maximum likelihood estimates, since the prior density is almost flat.  
\begin{table}
\centering
\scalebox{0.9}{
\begin{tabular}{|l|l|r|r|r|r|r|r|}
\hline
 $\lambda$ & & intercept   & age & gender &  experience    & wardtype & $\sigma^2$  \\
\hline
$\lambda=0.001$ & $\widehat{\beta}_{\BFI}$ (sd) & 0.522  (0.043) & 0.263  (0.034) & -0.502  (0.044) & -0.386  (0.034) & -0.011  (0.039) & 0.539  \\
& $\widehat{\beta}_{\com}$ (sd) &  0.332 (0.066) & 0.233 (0.052) & -0.503 (0.068) & -0.352 (0.052) & 0.075 (0.060) & 0.907 \\
\hline
$\lambda=0.1$ & $\widehat{\beta}_{\BFI}$ (sd) & 0.523 (0.043) & 0.264 (0.034) & -0.503 (0.044) & -0.386  (0.034) & -0.011  (0.039) & 0.537 \\
& $\widehat{\beta}_{\com}$ (sd) & 0.332 (0.066) & 0.233 (0.052) & -0.503 (0.068) & -0.352 (0.052) & 0.075 (0.060) & 0.907 \\
\hline
\end{tabular}}
\caption{The BFI estimates of the parameters in the linear regression model, $\widehat{\beta}_{\BFI}$, and the MAP estimates obtained from the analysis after combining the data, $\widehat{\beta}_{\com}$. The corresponding estimated standard deviations (sd) are given within the brackets.  The prior inverse covariance matrices are diagonal with the diagonal elements equal to  either $\lambda=0.001$ or $\lambda=0.1$. In the last column the estimates of $\sigma^2$, the variance of the error term, are given.  }
\label{Tab:est}
%\end{center}
\end{table}

The results are given in Table \ref{Tab:est}. It can be seen that the value of $\lambda$ hardly effects the estimates of the parameters; possibly because the total sample size is high. The BFI estimates for the regression parameters for the covariates age, gender and experience are similar to those obtained based on the combined data. For wardtype the estimates are close in absolute sense, but from the estimates and the relative large standard deviations it is clear that the contribution of this covariate to the model is minimal. 
The estimates of the intercept and the variance of the error term, $\sigma^2$, seem to differ substantially. This is possibly caused by the presence of heterogeneity across centers (e.g., varying hospital size and variances) for which is not corrected in the models (but will be in the next analysis). In the centers, the hospital size is taken into account via the intercept. This leads to different estimates of these intercepts across the centers and small variances of the error term. The BFI methodology combines the local estimates to a single estimate under the incorrect assumption of homogeneity, which leads to the differences of $\widehat{\bbeta}_{\BFI}$ and $\widehat{\bbeta}_{\com}$. In the next analysis, heterogeneity due to varying hospital sizes is taken into account and we will see that the differences between the estimates obtained with the two procedures will (almost) disappear. For the BFI methodology, but also if pooled data is available, it is important to correct for possible heterogeneity. We moreover leave out the covariate wardtype from the model.

Because the size of the hospital is predictive for the stress level, we want to add this variable to the model as well. This variable is a categorical variable with three categories (small, medium, large). For the combined data, the linear regression model that  includes the variable hospital size via category specific intercepts, is given by:
\begin{align*}
\mbox{stress}_{\ell i} ~=~ \beta_1 1_{\{z_\ell ~=~ \mbox{\footnotesize{small}}\}} ~+~ \beta_2 1_{\{z_\ell ~=~ \mbox{\footnotesize{medium}}\}} ~+~ \beta_3 1_{\{z_\ell ~=~ \mbox{\footnotesize{large}}\}} ~+~ \beta_4~ \mbox{age}_{\ell i}  ~+~ \beta_5~ \mbox{gender}_{\ell i} ~+~ \beta_6~ \mbox{exp}_{\ell i} ~+~ \varepsilon_{
\ell i},
\end{align*}
with $z_\ell$ the category of the hospital size in hospital $\ell$ (so small, medium or large) and $1_{\{z_\ell ~=~ \mbox{\footnotesize{small}}\}}$ is defined as 1 if hospital $\ell$ is small and zero otherwise. The functions $1_{\{z_\ell ~=~ \mbox{\footnotesize{medium}}\}}$ and $1_{\{z_\ell ~=~ \mbox{\footnotesize{large}}\}}$ are defined in a similar way. There is no general intercept in the model; this is hidden in the three intercepts. The model can be reformulated in a model that includes a general intercept (as was explained in Section \ref{sec: hetero}). To obtain a BFI aggregated model with category specific intercepts, we apply the BFI approach as described in Subsection \ref{subsec: clustering}.  The estimates are given in Table \ref{Tab:est2}. From the results we see that the estimates of the regression parameters obtained with the BFI methodology are very similar to those obtained based on the combined data; also for the three intercepts $\beta_1, \beta_2$ and $\beta_3$. However, there are still some differences between the estimates for $\sigma^2$, but these are  smaller than in the first analysis. Possibly more (unknown) variables need to be included in the model or there is heterogeneity in the variances across the centers. From the estimates of the intercepts, it is clear that there is a positive relationship between stress and the size of the hospital (adjusted for the other variables in the model): nurses in large hospitals seem to experience more stress than nurses in small hospitals. 
\begin{table}
\centering
\scalebox{0.9}{
\begin{tabular}{|l|l|r|r|r|r|r|r|r|}
\hline
 $\lambda$ & & I(small)  & I(medium) & I(large) & age & gender & experience  & $\sigma^2$ \\
\hline
$\lambda=0.001$ & $\widehat{\beta}_{\BFI}$ (sd) & 
0.004  (0.058) & 0.497  (0.043) & 0.958  (0.061) & 0.270  (0.035) & -0.478  (0.046) & -0.381  (0.036) & 0.581\\ 
& $\widehat{\beta}_{\com}$ (sd) & -0.024  (0.066) & 0.490  (0.063) & 0.917  (0.086) & 0.237  (0.049) &   -0.493  (0.064) & -0.352  (0.049) & 0.799  \\
\hline
$\lambda=0.1$ & $\widehat{\beta}_{\BFI}$ (sd) & 
0.004  (0.057) & 0.497  (0.043) & 0.958  (0.061) & 0.270  (0.035) &  -0.478  (0.046) & -0.381  (0.035) & 0.580 \\ 
& $\widehat{\beta}_{\com}$ (sd) & -0.024  (0.066) & 0.490  (0.063) &  0.916  (0.086) & 0.237  (0.049) &
  -0.492  (0.064) & -0.352  (0.049) & 0.799   \\
\hline
\end{tabular}}
\caption{  { The BFI estimates of the parameters in the linear regression model with a cluster effect for hospital size, $\widehat{\beta}_{\BFI}$, and the MAP estimates obtained from the analysis after combining the data, $\widehat{\beta}_{\com}$. The  estimated standard deviations (sd) are given within the brackets. The prior inverse covariance matrices are diagonal with the diagonal elements equal to } either $\lambda=0.001$ or $\lambda=0.1$. The abbreviations ``I(small)'', ``I(medium)'' and ``I(large)'' stand for the three intercepts for the categories small, medium, large. In the last column the estimates of $\sigma^2$, the variance of the error term, are given. }
\label{Tab:est2}
\end{table}

In a third analysis we include a hospital specific intercept in the model. Now, the variable hospital size is redundant as this effect is included in the hospital effect. The model for the combined data is given by:
\begin{align*}
\mbox{stress}_{\ell i} ~=~ \sum_{j=1}^{25} \beta_j 1_{\{\ell=j\}} ~+~ \beta_{26}~ \mbox{age}_{\ell i} ~+~ \beta_{27}~ \mbox{gender}_{\ell i} ~+~ \beta_{28}~ \mbox{exp}_{\ell i} ~+~   \varepsilon_{\ell i},
\end{align*}
with $1_{\{\ell=j\}}$ an indicator function defined as 1 if hospital $\ell$ is the $j^{th}$ hospital and zero otherwise. That means that for hospital $\ell$, $\sum_{j=1}^{25} \beta_j 1_{\{\ell=j\}} = \beta_\ell$. So, every hospital has its own specific intercept and there is no general intercept. We fit the model after merging the data and by combining the estimates in the different hospitals with the BFI methodology, as described in Subsection \ref{subsec: location specific outcome}. The results are given in Table \ref{Tab:est3}. Since the number of intercepts is large (for each hospital one intercept), we decided to leave out these estimates from the table, but made a scatter plot instead for comparison (not presented). The plot shows almost perfect agreement between the estimated intercepts based on the BFI methodology and the estimates found after combining the data. The estimates of the remaining parameters obtained with the two estimation procedures, shown in Table \ref{Tab:est3}, show nice agreement as well; also for the variance $\sigma^2$.  
\begin{table}
\centering
\scalebox{0.9}{
\begin{tabular}{|l|l|r|r|r|r|r|r|r|}
\hline
 $\lambda$ & & age & gender & experience  & $\sigma^2$ \\
\hline
$\lambda=0.001$ & $\widehat{\beta}_{\BFI}$ (sd) & 0.268  (0.036)  &  -0.452  (0.047) &   -0.364  (0.036) & 0.581  \\ 
& $\widehat{\beta}_{\com}$ (sd) & 
0.247  (0.043)  &  -0.474  (0.057) &   -0.357  (0.044) & 0.614\\ 
\hline
$\lambda=0.1$ & $\widehat{\beta}_{\BFI}$ (sd) & 0.268  (0.036) &  -0.452  (0.047) & -0.364  (0.036) &  0.580 \\
& $\widehat{\beta}_{\com}$ (sd) & 0.247 (0.043) &   -0.471  (0.057) & -0.357  (0.044) & 0.614 \\
\hline
\end{tabular}}
\caption{  { The BFI estimates of the parameters in the linear regression model with hospital specific intercepts, $\widehat{\beta}_{\BFI}$, and the MAP estimates obtained from the analysis after combining the data, $\widehat{\beta}_{\com}$. The corresponding estimated standard deviations (sd) are given within the brackets. The 25 estimated intercepts are not presented in the table. The prior inverse covariance matrices are diagonal with the diagonal elements equal to } either $\lambda=0.001$ or $\lambda=0.1$. In the last column the estimates of $\sigma^2$, the variance of the error term, are given. }
\label{Tab:est3}
\end{table}
 { Next, we consider the situation with  heterogeneity in the variance of the error term in the linear regression model. We allow center-specific intercepts and center-specific variances of the error term in the model. The estimates of the regression parameters hardly change (data not presented here). Taking into account heterogeneity across the centers can improve the results, but increases the number of model parameters that need to be estimated. }

\subsubsection{Prediction}
\label{subsec: prediction}
In the previous subsection we studied the performance of the BFI methodology for estimating the model parameters. In this subsection we focus on prediction.  

\bigskip

\begin{figure}
\begin{center}
\includegraphics[scale=0.95]{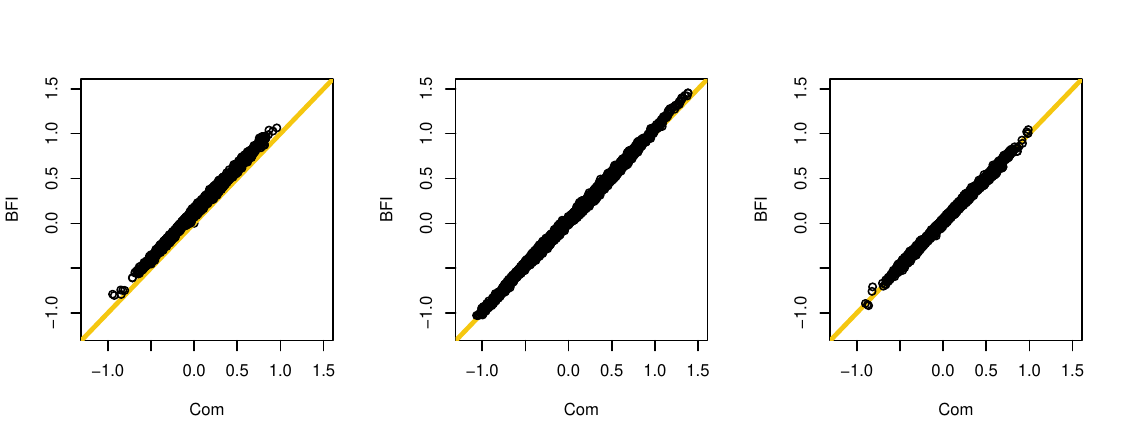}
\caption{Outcome predictions based on the BFI strategy (vertical axis) versus those based on the MAP estimates from the analysis obtained after combining the training data sets (horizontal axis). Left: Heterogeneous populations. Predictions are based on the model that includes the covariates age, gender, experience. Middle: Heterogeneous populations. Predictions are based on the model that includes the covariates hospital size, age, gender, experience. Right: Homogeneous populations. Predictions are based on the model that includes the covariates age, gender, experience. Perfect agreement corresponds to all points on the diagonal (yellow line). Here, $\lambda=0.1$. The plots look similar for other values of $\lambda$.
}
\label{Fig: scatterhetero}
\end{center}
\end{figure}

\noindent
{\bf Heterogenous populations}\\
\label{subsec:resultshetero}
To study the performance of a prediction model that has been estimated with the BFI strategy, we follow the steps:
\begin{enumerate}
\item In every hospital we randomly select the data of approximately 10\% of the nurses for the test-set. The remaining data form the training-set. The data of the nurses in this set will be used to estimate the BFI prediction model. The data in the test-set will be used to test the performance of the model. 
\item In every hospital we compute the MAP estimates of the model parameters based on the local data from the training sets only.
\item Based on the inference results from the hospitals, we compute the BFI estimates of the model parameters with the BFI methodology. 
\item Based on this estimated BFI model we predict the outcome (stress level) of the nurses in the test sets based on their covariate values. The prediction for the $i^{th}$ nurse in the $\ell^{th}$ hospital is denoted as $\widehat{Y}_{\BFI,\ell i}$.
\item Parallel to this, we merge all data from the training sets and fit the regression model by MAP estimation.
\item With this model we predict the outcomes of the nurses in the combined test data set based on their  covariate values. The predicted outcome for the $i^{th}$ nurse from hospital $\ell$ is denoted as $\widehat{Y}_{\com,\ell i}$.
\item We plot the points $(\widehat{Y}_{\com,\ell i},\widehat{Y}_{\BFI,\ell i})$ in a scatter plot.
\end{enumerate}
The steps above are repeated 50 times and all points are plotted in the same figure, see Figure \ref{Fig: scatterhetero} for three different settings. The predictions in the plot on the left were found based on the fitted model with the covariates age, gender, and experience. For the plot in the middle the covariate hospital size was included as well (as described in Subsection \ref{subsec: clustering}). In both cases $\lambda=0.1$. From the left plot we see that for the model that does not include the covariate hospital size, the BFI predictions are slightly higher than those found with the prediction model estimated based on the combined data. This is caused by the estimates of the intercept; in Table \ref{Tab:est} we already had seen that the intercept in the model fitted with the BFI method is higher than the estimated intercept in the model based on all data. This difference is due to heterogeneity of the data that is not taken into account in the model (see Subsection \ref{subsec: Data analysis} for a discussion). After adding the variable hospital size to the model this discrepancy disappears and there is a very strong agreement between the predictions obtained with the two methods. The variation in the predictions has increased which indicates a higher explained variance by the inclusion of the variable hospital size.  

\bigskip

\noindent
{\bf Homogeneous populations}\\
\label{subsec:resultshomo}
In the previous subsection we considered prediction accuracy of the BFI prediction model based on the data of nurses from the 25 hospitals. As mentioned before the nurses in the different hospitals may come from different populations. In this subsection we aim to study the performance of the BFI prediction model for homogeneous (nurse) populations. To be sure that the populations are homogeneous we randomize all nurses over the hospitals, keeping the sample sizes in the hospitals fixed. Now, the populations in the different hospitals can be seen as samples from the same population. Next, we follow the steps given in the previous subsection. This, including the randomization, is repeated 50 times. The variables we included in the model are age, gender, and experience. It can be seen that the agreement between the predictions is very strong. The scatter plot on the left in Figure \ref{Fig: scatterhetero} was obtained for the same model, but for the heterogeneous setting. In that case we saw some discrepancy between the predictions from the two models. Since this is not seen in the homogeneous setting and also not in the scatter plot for the models that take the hospital size into account, we conclude that the discrepancy was due to the heterogeneity that was not taken into account in the first simulation.

\section{Discussion}
\label{sec: Discussion}

In this paper we have extended the BFI methodology  for homogeneous to heterogeneous populations. The aim of the BFI methodology is to construct from the inference results obtained in multiple separate centers, what would have been found if the analysis had been performed on the combined data set. The key merit is that no individual data need to be transferred from the local centers to a central server. As a consequence, Data Transfer Agreement (DTA) for data sharing, can be simplified significantly. This may improve collaboration between researchers from different institutes and accelerate research. 

In the BFI framework, statistical models are fitted in the separate centers based on local data only. So, in every center someone with sufficient knowledge of statistics and R needs to be available to do the analysis. Of course, the statistician who is concerned with combining the separate inference results can assist and can even provide code to be sure that the analyses in the separate centers are consistent.  { It is therefore important that a single communication with the local centers is sufficient to calculate the BFI estimators.  }  

 { For different types of heterogeneity, new expressions of the BFI estimators have been derived. Asymptotically, the BFI estimators have been proven to be efficient (minimum variance) and we show that no information is lost as a consequence of the fact that the data cannot be combined. Simulation studies have shown that the performance of the BFI estimator is also good for finite samples, and better than that of the weighted average estimator.} Furthermore, in this paper it is explained how to do the analyses in R with the software package {\tt BFI} that we developed to make the methodology easily accessible for the user. The mathematical details are given in two appendices, and can be ignored if one's interest is solely in the application of BFI.   

 { It may happen that communication between the central server and some data centers is intermittent or delayed. In that case, the BFI estimators can be calculated based on the estimation results available so far. As soon as more centers have sent their results, the BFI estimators can be recalculated, including the results from the delayed centers. This can be easily done by using the expression of the BFI estimators. In many other federated analysis methods, in contrast, estimates are found by cycling around the centers and updating parameter estimates based on the local data. Then, if one or more centers are included in the estimation process at a later moment, the entire optimization process needs to be repeated, which can be a rather time-consuming process.}

% { If centers are located in different regions, the populations in the centers may differ. That means that the distributions of the covariates may be different across the centers. If the parameters $\btheta_1$ for the regression coefficients are assumed to be independent of $\btheta_2$ for the covariate distributions, this will not affect the BFI estimate of $\btheta_1$.  }  This means that heterogeneity across the population characteristics in the different centers do not affect the estimates that describe the association between the predictors and the outcome, under the assumption that the regression parameters are equal in the centers.  { This was also seen in the simulation studies.} %So, even if the centers are located in different parts of the country or the world and serve populations with different characteristics, the strategy of combining the inference results for the regression models still holds under the assumption that the regression parameters are equal in the centers. %And even if the latter is not expected for some covariates, the BFI strategy can take this into account.

 { The prior of the parameters is taken equal to a zero-mean Gaussian distribution. This assumption allows the derivation of explicit expressions for the BFI estimators. For other prior distributions this may not be the case. If a Gaussian prior is not suitable for a parameter, for example because it is positive by definition, it can be transformed (e.g., via a log transformation). For example, for the variance of the error term in a linear regression model, the Gaussian prior for the log transformation of the parameter is used and implemented in the R package. The Gaussian prior corresponds to a ridge penalty, which is often used in practice to reduce overfitting. If one also wants to do selection, a lasso penalty is more common and a different prior distribution must be assumed. Then the BFI estimates must be found by numerical optimization. }

 { For the centers different covariance matrices for the Gaussian prior may be chosen. One reason to do this could be the local sample size. The smaller the variance of the Gaussian prior, the more the estimates are shrunken to zero. Also if there is a difference in reliability of the data across the centers (data in some centers are ``cleaner'' than in others), different prior covariances can be used. It is up to the user to decide whether to assume equal priors or not.}

The sets of variables available for fitting a regression model may differ across the centers. This happens, for instance, if some patients' or individuals' characteristics are measured and documented in most centers, but not in all. If a missing variable may be predictive for the outcome, a single or multiple regression method can be applied to impute the missing values\cite{Buuren}. Then, a regression model with this missing variable as an outcome and the original outcome variable and the remaining variables as covariates is fitted, by applying the BFI approach in the centers in which this ``missing variable'' has been measured. Next, this estimated regression model is used to predict the variable values in the center in which the variable was not measured. After a single or a multiple imputation, the BFI strategy as described before can be used.  

 { The BFI estimators are defined as the maximizers of an approximation of the log posterior density (second order Taylor expansions) for the merged data set. In the homogeneous case, these approximations are known to be accurate if the total sample size is sufficiently large (compared to the dimension of parameter space). However, if parameters are assumed to be distinct across centers, the local sample sizes need to be sufficiently large as well. If the total or local sample sizes are small, a higher order approximation (third or higher order of the Taylor expansion) should yield more accurate results. This will be studied in a new project in which we want to consider overfitting in high-dimensional models. The same holds for the BFI estimates of the asymptotic covariance matrix and, thus, for the standard deviations. }

The theory for the BFI approach has been developed for parametric models, including generalized linear models (GLMs) and survival models.\cite{Pazira}   { The methodology has been tested for different data sets, in this paper for the heterogeneous setting and for the homogeneous setting for GLMs in Jonker et al\cite{Jonker} and for parametric survival models in Pazira et al\cite{Pazira}.} 
The Cox model for time-to-event data is a semi-parametric model. Developing the theory for this model and for even more complex models, will be a next step. The development of the theory will be accompanied by updates of the R package BFI.\cite{Pazira_package}

The BFI methodology makes it possible to obtain the statistical power of the combined data set {\em without actually combining the data}. DTA's can hence be simplified and collaboration between centers may increase.   

\bigskip

\noindent
{\bf{Software}}\\
 { The R package BFI and a detailed manual are available on CRAN: {\tt https://CRAN.R-project.org/package=BFI}. More information can also be found on the webpage: {\tt https://hassanpazira.github.io/BFI/}.}

\bigskip

\noindent
{\bf{Data availability statement}}\\
The data are available in the R package BFI. 

\bigskip

\noindent
{\bf{Funding}}\\
This research was supported by an unrestricted grant of Stichting Hanarth Fonds, The Netherlands. 

%\bigskip

%\noindent
%{\bf{Acknowledgement}}\\
% { We would like to thank the anonymous reviewers and associate editor for their valuable comments.}

\bigskip

\noindent
{\bf{Conflict of interest statement}}\\
We have no conflicts of interest to disclose. 

\bigskip

\noindent
{\bf{Ethics statement}}\\
This research is solely aimed at advancing knowledge in the field of federated inference. No sensitive data is involved. Therefore, ethical considerations are not applicable for this research.

\bigskip

\noindent
{\bf{Orcid}}\\
Marianne Jonker: https://orcid.org/0000-0003-0134-8482 

%\bibliographystyle{apalike} %apalike %plain
%\bibliographystyle{natbib} %apalike %plain
%\bibliography{Bibfor-BFI-GLM} 

%\end{document}

\section*{Appendix I: Bayesian Federated Inference in R}
We have written the software package {\tt BFI} in R for doing the BFI calculations\cite{Pazira_package}. Here we explain how to do BFI analyses.

\subsection*{MAP estimation}
To compute the MAP estimates of the parameters in a regression model, the command {\tt MAP.estimation} can be used. To apply this command, the data has to be in a specific form and the inverse covariance matrix of the Gaussian prior needs to be chosen. The analysis below is for the combined data set {\tt Nurse}. The estimates in the separate hospitals can be obtained with the same commands, but with the local data sets instead.   
\begin{verbatim}
   library(BFI)
   M      <- data.frame(age=Nurses$age,gender=Nurses$gender,exp=Nurses$experien,
                        wardtype=Nurses$wardtype)
   Lambda <- inv.prior.cov(M,lambda=0.01,family="gaussian")
   fit    <- MAP.estimation(Nurses$stress,X=M,family="gaussian",Lambda)
\end{verbatim}  
The command {\tt inv.prior.cov} creates a diagonal inverse covariance matrix for the prior distribution of the correct dimension. Based on the characteristics of the covariates (continuous or categorical) in {\tt M} and the number of nuisance parameters, the number of model parameters is computed (the number of regression parameters for a categorical variable equals the number of levels minus one and a linear model ({\tt family="gaussian"}) has one nuisance parameter, the variance of the measurement error). The argument {\tt lambda=0.01} means that all elements on the diagonal of $\Lambda$ equal 0.01. %If two values are given, e.g., {\tt lambda=c(0.1,0.01)}, the first value (in this case 0.1) is on the diagonal of the inverse covariance matrix for all regression parameters, whereas the last value (0.01 here) is the value for the standard deviation of the measurement error. The latter is only valid for a linear regression model. Note that the values assigned to {\tt lambda} should be positive always, as they equal 1 divided by a variance. Because $\Lambda$ is the inverse diagonal covariance matrix, values of $\lambda$ close to zero indicate a large variance of the prior distribution (the prior is not very informative) and a large value indicates a small variance (a very informative prior). Furthermore, because $\Lambda$ is a diagonal matrix, independence between the parameters is assumed. 
%After setting the inverse covariance matrix, the MAP estimates can be computed with the function {\tt MAP.estimation}. This command maximizes the log posterior density:
%\begin{verbatim}
%   fit <- MAP.estimation(Nurses$stress,X=M,family="gaussian",Lambda)
%\end{verbatim}
The arguments of {\tt MAP.estimation} are the outcome variable {\tt Nurses\$stress}, the covariate data in  {\tt M}, the type of the model ({\tt family="gaussian"}) and the matrix {\tt Lambda}. %The MAP estimates are obtained by numerical maximization with the so-called Limited-memory Broyden-Fletcher-Goldfarb-Shanno algorithm with Bound Constraints (L-BFGS-B), \cite{Byrd}. As a starting point of the maximization 0 is taken by default, the mean of the prior distribution, but a different starting point can be chosen. 
The inference results are stored in the list {\tt fit}. A summary of it can be found with {\tt summary(fit)}.  

When applying the BFI approach, the analyses are performed in every hospital and the results in {\tt fit} are sent to the central server. There, the results from the different hospitals are combined. This is explained below.

\subsection*{BFI for homogeneous populations}

Suppose that all hospitals have sent their output to the central server. For ease of notation, we assume these outputs are stored in {\tt fit1, fit2, \ldots, fit25}. From every output the relevant elements need to be selected and combined. %In case of three hospitals the simplest way to do this is:
% \begin{verbatim}
%     thetahats <- list(fit1$theta_hat,fit2$theta_hat,fit3$theta_hat)
%     Ahats     <- list(fit1$A_hat,fit2$A_hat,fit3$A_hat)  
%     priors    <- list(fit1$Lambda,fit2$Lambda,fit3$Lambda)
% \end{verbatim}
% It is straightforward to extend this to more hospitals. Now the R code for computing the BFI estimators can be applied:
% \begin{verbatim}
%     priors <- append(priors,list(Lambda))
%     fitbfi_homo <- bfi(thetahats,Ahats,priors)
%     summary(fitbfi_homo)
% \end{verbatim}
%
%\noindent
 { If the number of hospitals is high, it is easier to work with a for-loop. With the following code, all relevant elements of 25 local centers are created and combined by the main function {\tt bfi()}:}
\begin{verbatim}
    Ms <- fits <- thetahats <- Ahats <- priors <- Lambdas <- list()
    for (l in 1:25) {
      Ms[[l]] <- data.frame(age = Nurses$age[Nurses$hospital==l],
                            gender = Nurses$gender[Nurses$hospital==l],
                            exp = Nurses$experien[Nurses$hospital==l],
                            wardtype = Nurses$wardtype[Nurses$hospital==l])
      Lambdas[[l]] <- inv.prior.cov(Ms[[l]], lambda=0.01, family="gaussian")
      fits[[l]] <- MAP.estimation(y=Nurses$stress[Nurses$hospital==l], X=Ms[[l]],
                                  family="gaussian", Lambda=Lambdas[[l]])
      thetahats[[l]] <- fits[[l]]$theta_hat
      Ahats[[l]]     <- fits[[l]]$A_hat
      priors[[l]]    <- fits[[l]]$Lambda
    }
    priors_all <- append(priors,list(Lambda))
    fitbfi_homo <- bfi(theta_hats=thetahats, A_hats=Ahats, Lambda=priors_all)
    summary(fitbfi_homo)
\end{verbatim}

Here {\tt Lambda} is the inverse covariance matrix of the prior for the (fictive) combined data. The command {\tt bfi} combines the estimates from the different hospitals into the BFI estimates. The outcome {\tt fitbfi\_homo} is a list with the BFI estimates $\widehat{\btheta}_\BFI$ and $\widehat{\bA}_\BFI$. The command {\tt summary(fitbfi\_homo)} gives the BFI estimates (and more information).

\subsection*{BFI for heterogeneous populations}
Different types of heterogeneity have been discussed in Section \ref{sec: hetero}. Below we will explain how to do the analyses in R. 

\bigskip

\noindent
{\underline{Heterogeneity of population characteristics}}\\
Heterogeneity across population characteristics in the centers implies that the value of the parameter $\btheta_2$ differs across centers. Because the bfi-command estimates the parameter $\btheta_1$ (and its curvature matrix $\widehat{A}_1$), and these estimates are not affected by $\btheta_2$, the R-code explained in the previous subsection can still be applied.    

\bigskip

\noindent
{\underline{Heterogeneity across outcome means}}\\
Suppose the intercepts  differ across hospitals. To take this variation into account we allow a hospital specific intercept in the regression model. Instead of one general intercept there are $L=25$ intercepts; an increase of $L-1$ parameters. The dimension of the inverse covariance matrix {\tt Lambda} for the fictive combined data set changes as well. For a diagonal matrix with 0.01 at the diagonal, this matrix can be obtained by 
\begin{verbatim}
   Lambda <- inv.prior.cov(M,lambda=0.01,stratified=TRUE,strat_par=1,L=25)
   priors_all <- append(priors,list(Lambda))
\end{verbatim}
These commands replace the two corresponding commands above. The argument {\tt L=25} has to be added to indicate the number of centers, and thus the number of location specific intercepts. This matrix should be appended to the list {\tt priors} instead. The MAP estimates can be obtained with the command {\tt bfi}, but it needs to be made explicit that the hospitals may have different intercepts: 
\begin{verbatim}
   fitbfi_hetero <- bfi(thetahats,Ahats,priors_all,stratified=TRUE,strat_par=1)
\end{verbatim}
For this stratified analysis extra arguments have been added: {\tt stratified=TRUE} and \verb|strat_par=1|. The first argument indicates that the full model stratifies with respect to the different hospitals. The default is {\tt stratified=FALSE}. If \verb|strat_par=1| there is stratification with respect to the intercept and if \verb|strat_par=2| this is the case for the variance of the measurement error in a linear regression model. A summary of the results can be obtained by \verb |summary(fitbfi_hetero)|. This gives a list with estimates, starting with the hospital specific intercepts.
  
\bigskip

\noindent
{\underline{Heterogeneity due to clustering}}\\
An example of a cluster variable is hospital size. For all nurses in a hospital this covariate is constant and, as a consequence, the effect of hospital size on stress cannot be estimated within a hospital. However, the model for the (fictive) combined data could include this covariate if there is variation across the hospitals (Subsection \ref{subsec: clustering}). Here we explain how to do the analyses in R. In practice, every local hospital sends its size (small, medium, large) to the central server. Then, a vector with all sizes is defined in R. Suppose this vector is named {\tt Hsize}. After fitting the local models (like explained before), the estimated model for the (fictive) combined data can be obtained with:
\begin{verbatim}
   Hsize <- c() 
   for (i in 1:25)
        Hsize[i] <- Nurses[Nurses$hospital==i,]$hospsize[1]
   LambdaCom <- inv.prior.cov(Mi,lambda=0.01,stratified=TRUE,center_spec=Hsize,L=25)
   priors_all <- append(priors,list(LambdaCom))
   fitbfi_hetero <- bfi(thetahats,Ahats,priors_all,stratified=TRUE,center_spec=Hsize)
   summary(fitbfi_hetero)
\end{verbatim}
The commands return a list with categorical specific intercepts and the estimates of the remaining parameters. 
 
%For a continuous covariate the analysis is more complex. First the BFI estimates are computed allowing for different intercepts across the hospitals. Next, the estimated intercepts are plotted against the continuous hospital sizes. A least square line is fitted through these points with the command {\tt lm} to obtain estimates of the parameters $\nu_0$ and $\nu_1$ in the regression model (\ref{BFI cat model}).
%\begin{verbatim}
%    fitbfi <- bfi(thetahats,Ahats,priors,stratified=T,strat_par=1)  
%    plot(Hsize,fitbfi$thetahats[1:25],xlab="hospital size",ylab="intercepts")
%    lm(fitbfi$thetahats[1:25],Hsize)
%\end{verbatim}

\section*{Appendix II: Mathematical derivations of the BFI estimators}
In this appendix the mathematical derivations of the BFI estimators are given for three settings: 
\begin{enumerate}
\item[] Appendix II.A: Homogeneity across centers. 
\item[] Appendix II.B: Heterogeneity across centers, center-specific parameter, e.g., the intercept.
\item[] Appendix II.C: Heterogeneity across centers,  { due to clustering, e.g.\ geospatial regions. }
\end{enumerate}

\subsection*{Appendix II.A: Homogeneity across centers}
\label{subsec: Appendix A}

 { In this appendix we derive expressions of the BFI estimators under the assumption that the variables $(\bX_{\ell i},Y_{\ell i}), i=1,\ldots,n_\ell, \ell=1,\ldots, L$ are independent and identically distributed.} In equations (\ref{eq:logpost}) and (\ref{eq:loppostl}) in Section \ref{sec: BFI} we have seen that the log posterior densities for the (fictive) combined data set $\bD$ and for the subset $\bD_\ell$ equal 
\begin{align}
\log \big\{ p(\btheta|\bD) \big\} 
&=\; \log \big\{p(\theta_1)\big\} \;+\; \log \big\{p(\theta_2)\big\} \;+\; \sum_{\ell=1}^L \sum_{i=1}^{n_\ell}\log \big\{p(y_{\ell i}| \bx_{\ell i},\btheta_1)\big\}  \;+\; \sum_{\ell=1}^L \sum_{i=1}^{n_\ell}\log \big\{ p(\bx_{\ell i}|\btheta_2)\big\} \;-\; \log \big\{p(\bD)\big\} \label{eq:logpostA}\\
\log \big\{p_\ell(\btheta|\bD_\ell)\big\} &=\; \log \big\{p_\ell(\theta_1)\big\} \;+\; \log \big\{p_\ell(\theta_2)\big\} \;+\; \sum_{i=1}^{n_\ell}\log \big\{p(y_{\ell i}| \bx_{\ell i},\btheta_1)\big\} \;+\; \sum_{i=1}^{n_\ell}\log \big\{p(\bx_{\ell i}|\btheta_2)\big\} \;-\; \log \big\{p_\ell(\bD_\ell)\big\}.
\label{eq:logpostlA}
\end{align}
By reordering the terms in equation  (\ref{eq:logpostlA}), it follows that for every center $\ell$
\begin{align*}
\sum_{i=1}^{n_\ell}\log \big\{p(y_{\ell i}| \bx_{\ell i},\btheta_1)\big\} \;+\;  \sum_{i=1}^{n_\ell}\log \big\{p(\bx_{\ell i}|\btheta_2)\big\} \;=\; \log \big\{p_\ell(\btheta|\bD_\ell)\big\} - \log \big\{p_\ell(\btheta_1)\big\} \;-\; \log \big\{p_\ell(\btheta_2)\big\} + \log \big\{p_\ell(\bD_\ell)\big\}.  
\end{align*}
Next, summing over all centers yields
\begin{align}
\label{eq:sum}
\lefteqn{\sum_{\ell=1}^L\sum_{i=1}^{n_\ell}\log \big\{p(y_{\ell i}| \bx_{\ell i},\btheta_1)\big\} \;+\;  \sum_{\ell=1}^L\sum_{i=1}^{n_\ell}\log \big\{p(\bx_{\ell i}|\btheta_2)\big\}} \nonumber\\
&~~~~~~~~ =\; \sum_{\ell=1}^L\log \big\{p_\ell(\btheta|\bD_\ell)\big\} - \log \Big\{\prod_{\ell=1}^L p_\ell(\btheta_1)\Big\} \;-\; \log \Big\{\prod_{\ell=1}^L p_\ell(\btheta_2)\Big\} \;+\; \log \Big\{\prod_{\ell=1}^L p_\ell(\bD_\ell)\Big\}.  
\end{align}
By inserting the right hand side of equation (\ref{eq:sum}) into the right hand side of equation (\ref{eq:logpostA}) this yields  
\begin{align}
\label{eq:full_into_subsets}
\log \big\{p(\btheta|\bD)\big\} \;=\; \sum_{\ell=1}^L \log \big\{p_\ell(\btheta|\bD_\ell)\big\}+\log\Bigg\{\frac{p(\btheta_1)}{\prod_{\ell=1}^L p_\ell(\btheta_1)}\Bigg\}+\log\Bigg\{\frac{p(\btheta_2)}{\prod_{\ell=1}^L p_\ell(\btheta_2)}\Bigg\}-\log\Bigg\{\frac{ p(\bD)}{\prod_{\ell=1}^L p_\ell(\bD_\ell)}\Bigg\}. 
\end{align}

We expressed the log posterior densities of the combined data, $\log \{p(\btheta|\bD)\}$, in terms of the log posterior densities of the local data sets, $\log \{p_\ell(\btheta|\bD_\ell)\}$. However, the final aim is to express the MAP estimator $\widehat{\btheta}$ based on the (fictive) combined data set $\bD$ in terms of the MAP estimators based on the local data sets $\bD_\ell$. This will be done next. We approximate the log posterior densities for the data set $\bD_\ell$ by a Taylor expansion up to the quadratic order in $\btheta$ around the MAP estimator $\widehat{\btheta}_\ell$:
\begin{align*}
\log \big\{p_\ell\big(\btheta|\bD_\ell\big)\big\} \;=\; \log \big\{p_\ell\big(\widehat{\btheta}_\ell|\bD_\ell\big)\big\} -\tfrac{1}{2}\big(\btheta-\widehat{\btheta}_\ell\big)^t \widehat{\bA}_\ell(\btheta-\widehat{\btheta}_\ell\big) + O_p\big(\|\widehat{\btheta}_\ell-\btheta\|^3\big),
\end{align*}
with $\widehat{\bA}_\ell$ equal to minus the second derivative of $\log \{p_\ell(\btheta|\bD_\ell)\}$ with respect to $\btheta$, evaluated at $\widehat{\btheta}_\ell$. The linear term in the Taylor expansion is equal to zero and therefore missing in the expansion; the MAP estimator maximizes the log posterior density and the first derivative evaluated at $\widehat{\btheta}_\ell$ is therefore equal to zero. The last term in the Taylor expansion is equal to $O_p(\|\widehat{\btheta}_\ell-\btheta\|^3)=\|\widehat{\btheta}_\ell-\btheta\|^3 O_p(1)$, where $O_p(1)$ represents a term that is bounded in probability for the sample size going to infinity \cite{vdVaart}. For $\btheta$ in a small neighborhood of $\widehat{\btheta}_\ell$, the term $\|\widehat{\btheta}_\ell-\btheta\|^3$ will be close to zero (in probability), and the remainder term $O_p(\|\widehat{\btheta}_\ell-\btheta\|^3)$ is small compared to the other terms in the Taylor expansion which are of an order of at most $\|\widehat{\btheta}_\ell-\btheta\|^2$. 

Since the log posterior density in equation
(\ref{eq:logpostlA}) is decomposed in terms that depend on either $\btheta_1$ or $\btheta_2$, but never on both, the matrices $\widehat{\bA}_\ell, \ell=1,\ldots,L$ are diagonal block matrices: 
\begin{align}
\widehat{\bA}_\ell=\left(\!\begin{array}{cc}
\widehat{\bA}_{1,\ell} & {\bf 0}\\
{\bf 0} & \widehat{\bA}_{2,\ell}\end{array}\!
\right), \nonumber
\end{align}
with the blocks $\widehat{\bA}_{1,\ell}$ and $\widehat{\bA}_{2,\ell}$ equal to minus the second derivative matrices for $\btheta_1$ and  $\btheta_2$, respectively, and the log posterior densities can be approximated by
\begin{align*}
%\log \big\{p\big(\btheta|\bD\big)\big\} &\approx\; \log \big\{p\big(\widehat{\btheta}|\bD\big)\big\} \;-\; \tfrac{1}{2}\big(\btheta_1-\widehat{\btheta}_1\big)^t\widehat{\bA}_1\big(\btheta_1-\widehat{\btheta}_1\big) \;-\; \tfrac{1}{2}\big(\btheta_2-\widehat{\btheta}_2\big)^t\widehat{\bA}_2\big(\btheta_2-\widehat{\btheta}_2\big)\\[8pt]
\log \big\{p_\ell\big(\btheta|\bD_\ell\big)\big\} &=\; \log \big\{p_\ell\big(\widehat{\btheta}_\ell|\bD_\ell\big)\big\} \;-\; \tfrac{1}{2}\big(\btheta_1-\widehat{\btheta}_{1,\ell}\big)^t\widehat{\bA}_{1,\ell}\big(\btheta_1-\widehat{\btheta}_{1,\ell}\big) \;-\; \tfrac{1}{2}\big(\btheta_2-\widehat{\btheta}_{2,\ell}\big)^t\widehat{\bA}_{2,\ell}\big(\btheta_2-\widehat{\btheta}_{2,\ell}\big) + O_p\big(\|\widehat{\btheta}_\ell-\btheta\|^3\big).
\end{align*}
By substituting this expansion for $\log\{ p_\ell(\btheta|\bD_\ell)\},\ell=1,\ldots,L$, into the relation (\ref{eq:full_into_subsets}), we obtain:
\begin{align}
\log \big\{p(\btheta|\bD)\big\} &=\; -\tfrac{1}{2}\sum_{\ell=1}^L \big(\btheta_1-\widehat{\btheta}_{1,\ell}\big)^t\widehat{\bA}_{1,\ell}\big(\btheta_1-\widehat{\btheta}_{1,\ell}\big) -\; \tfrac{1}{2}\sum_{\ell=1}^L\big(\btheta_2-\widehat{\btheta}_{2,\ell}\big)^t\widehat{\bA}_{2,\ell}\big(\btheta_2-\widehat{\btheta}_{2,\ell}\big) \nonumber\\
&\; ~~~~~~ 
\;+\;\log\Bigg\{\frac{p(\btheta_1)}{\prod_{\ell=1}^L p_\ell(\btheta_1)}\Bigg\}
\;+\;\log\Bigg\{\frac{p(\btheta_2)}{\prod_{\ell=1}^L p_\ell(\btheta_2)}\Bigg\} \;+\; B \;+\; O_p\Big(\sum_{\ell=1}^L\|\widehat{\btheta}_\ell-\btheta\|^3\Big).  
\label{eq:logp}
\end{align}
where $B$ is a term that depends on the data, but is not a function of $\btheta=(\btheta_1,\btheta_2)$.
%Seen as a function of $\btheta_1$ and $\btheta_2$, this yields
%\begin{align}
%\big(\btheta_1-\widehat{\btheta}_1\big)^t \widehat{\bA}_1\big(\btheta_1-\widehat{\btheta}_1\big) 
%&\approx\; \sum_{\ell=1}^L \big(\btheta_1-\widehat{\btheta}_{1,\ell}\big)^t\widehat{\bA}_{1,\ell}\big(\btheta_1-\widehat{\btheta}_{1,\ell}\big)  \;-\; 2\log\Bigg\{\frac{p(\btheta_1)}{\prod_{\ell=1}^L p_\ell(\btheta_1)}\Bigg\}
%\;+\; B_1. 
%\label{eq:theta1} \\
%\big(\btheta_2-\widehat{\btheta}_2\big)^t \widehat{\bA}_2\big(\btheta_2-\widehat{\btheta}_2\big) 
%&\approx\; \sum_{\ell=1}^L \big(\btheta_2-\widehat{\btheta}_{2,\ell}\big)^t\widehat{\bA}_{2,\ell}\big(\btheta_2-\widehat{\btheta}_{2,\ell}\big)  \;-\; 2\log\Bigg\{\frac{p(\btheta_2)}{\prod_{\ell=1}^L p_\ell(\btheta_2)}\Bigg\}
%\;+\; B_2. 
%\label{eq:theta2} 
%\end{align}
%for $B_1$ and $B_2$ terms that depend on the data, but are no functions of $\btheta_1$ and $\btheta_2$. 
Now choose the prior densities $\btheta_1 \rightarrow p(\btheta_1)$ and $ \btheta_2 \rightarrow p(\btheta_2)$ in the combined data set and $\btheta_1 \rightarrow p_\ell(\btheta_1)$ and $\btheta_2 \rightarrow p_\ell(\btheta_2)$ in center $\ell$ to be Gaussian with mean zero and inverse covariance matrices $\bLambda_1$ and $\bLambda_2$ in the combined data set, and $\bLambda_{1,\ell}$ and $\bLambda_{2,\ell}$ in center $\ell$: e.g., $p(\btheta_1) = (\det \bLambda_1 / (2\pi)^d)^{1/2}\exp(-\tfrac{1}{2}\btheta_1^t\bLambda_1\btheta_1)$. Inserting the expressions of the densities into (\ref{eq:logp}) yields
\begin{align}
\label{eq:1}
\log \big\{p(\btheta|\bD)\big\} &=\; -\tfrac{1}{2}\sum_{\ell=1}^L \big(\btheta_1-\widehat{\btheta}_{1,\ell}\big)^t\widehat{\bA}_{1,\ell}\big(\btheta_1-\widehat{\btheta}_{1,\ell}\big)  \;-\; \tfrac{1}{2} \sum_{\ell=1}^L \big(\btheta_2-\widehat{\btheta}_{2,\ell}\big)^t\widehat{\bA}_{2,\ell}\big(\btheta_2-\widehat{\btheta}_{2,\ell}\big)\nonumber\\
&~~~~~~\;-\tfrac{1}{2} \btheta_1^t \Big(\Lambda_1-\sum_{\ell=1}^L \bLambda_{1,\ell}\Big)\btheta_1 \;-\tfrac{1}{2} \btheta_2^t \Big(\Lambda_2-\sum_{\ell=1}^L \bLambda_{2,\ell}\Big)\btheta_2 
\;+\; B'~\;+\; O_p\Big(\sum_{\ell=1}^L\|\widehat{\btheta}_\ell-\btheta\|^3\Big) \nonumber\\
&=: \; \Omega_{\BFI}(\theta) \;+\; O_p\Big(\sum_{\ell=1}^L\|\widehat{\btheta}_\ell-\btheta\|^3\Big), 
\end{align}
for $B'$ a term that depends on the data, but not of $\btheta_1$ and $\btheta_2$.
The function $\btheta \rightarrow \Omega_{\BFI}(\btheta)$ in equation (\ref{eq:1}) is quadratic function of $\btheta_1$ and $\btheta_2$. Maximizing $\btheta \rightarrow \Omega_{\BFI}(\btheta)$ with respect to $\btheta=(\btheta_1,\btheta_2)$ yields the BFI estimators
\begin{align*}
\hspace*{-5mm} & \widehat{\btheta}_{1,\BFI} \;:=\;  \big(\widehat{\bA}_{1,\BFI}\big)^{-1}\sum_{\ell=1}^L \widehat{\bA}_{1,\ell}\widehat{\btheta}_{1,\ell},
~~~~~~~~~~~~~\widehat{\bA}_{1,\BFI} := \sum_{\ell=1}^L \widehat{\bA}_{1,\ell}+\bLambda_1-\sum_{\ell=1}^L \bLambda_{1,\ell},
%\label{eq:recover_theta1}
\\
\hspace*{-5mm} & \widehat{\btheta}_{2,\BFI} \;:=\; \big(\widehat{\bA}_{2,\BFI}\big)^{-1}\sum_{\ell=1}^L \widehat{\bA}_{2,\ell}\widehat{\btheta}_{2,\ell}, ~~~~~~~~~~~~~\widehat{\bA}_{2,\BFI} \;:=\; \sum_{\ell=1}^L \widehat{\bA}_{2,\ell}+\bLambda_2-\sum_{\ell=1}^L \bLambda_{2,\ell},
%\label{eq:recover_theta2}
\end{align*}
where $\widehat{\bA}_{1,\BFI}$ and $\widehat{\bA}_{2,\BFI}$ equal minus the second derivative of $\Omega_{\BFI}$ with respect to $\btheta_1$ and $\btheta_2$. In Appendix III.B the asymptotic distribution of the BFI estimators are derived.

\subsection*{Appendix II.B: Heterogeneity across centers, center-specific parameter}
\label{subsec: Appendix B}

Suppose that the vector of regression parameters can be subdivided into two parts. One part is equal across the centers and the other part may vary. A special case is the situation in which the intercepts vary.  { In the calculations of the BFI estimator, we assume that the covariates are statistically independent between the individuals within and across the centers. We, moreover, assume that the outcome variables given the covariates and the center are independent.}

Suppose that the vector $\btheta$ can be decomposed as $\btheta=(\btheta_1,\btheta_2)=(\btheta_{1a},\btheta_{1b},\btheta_2)$, where, as before, $\btheta_2$ is the vector of parameters that specifies the distribution of the covariates. The parameter $\btheta_1=(\btheta_{1a},\btheta_{1b})$ is decomposed so that $\btheta_{1a}$ is the vector of (regression) parameters that is assumed to be equal across the centers, and $\btheta_{1b}$ is the vector of (regression) parameters that may vary. In this appendix we consider the situation in which every center has its own specific vector of parameters: $\btheta_{1b,1}, \ldots, \btheta_{1b,L}$ for the $L$ centers.  
The vector of parameters in the combined data set is equal to $\btheta=(\btheta_{1a},\btheta_{1b,1},\ldots,\btheta_{1b,L},\btheta_{2})$, where $\btheta_{1b,\ell}$ is the parameter vector in center $\ell$. If only the intercepts vary across the centers, $\btheta_{1b,\ell}$ is one-dimensional, but for now we allow $\btheta_{1b,\ell}$ to be a vector. 

For simplicity of notation we assume that $\btheta_{1a}, \btheta_{1b}$ and $\btheta_2$ are independent:  $p(\btheta)=p(\btheta_{1a})p(\btheta_{2})\prod_{\ell=1}^L p(\btheta_{1b,\ell})$ for the combined data set, and in center $\ell$: $p_\ell(\btheta_{1a},\btheta_{1b,\ell},\btheta_{2})=p_\ell(\btheta_{1a})p_\ell(\btheta_{1b,\ell})p_\ell(\btheta_{2})$. As before, the log posterior densities can be written as 
\begin{align*}
\lefteqn{\log \big\{p(\btheta|\bD)\big\}}\\
&=\; \log \big\{p(\btheta_{1a})\big\} + 
\sum_{\ell=1}^L \log \big\{p(\btheta_{1b,\ell})\big\} + \log \big\{p(\btheta_{2})\big\} + \sum_{\ell=1}^L\sum_{i=1}^{n_\ell} \log \big\{p(y_{\ell i}|\btheta_{1a},\btheta_{1b,\ell},\bx_{\ell i})\big\}  + \sum_{\ell=1}^L\sum_{i=1}^{n_\ell} \log \big\{p(\bx_{\ell i}|\btheta_{2})\big\} -\log \big\{p(\bD)\big\},
\end{align*}
and 
\begin{align*}
\lefteqn{\log \big\{p_\ell(\btheta_{1a},\btheta_{1b,\ell},\btheta_2|\bD_\ell)\big\}}\\
&=\; \log \big\{p_\ell(\btheta_{1a})\big\} + \log \big\{p_\ell(\btheta_{1b,\ell})\big\} + \log \big\{p_\ell(\btheta_{2})\big\} + \sum_{i=1}^{n_\ell} \log \big\{p(y_{\ell i}|\btheta_{1a},\btheta_{1b,\ell},\bx_{\ell i})\big\} + \sum_{i=1}^{n_\ell} \log \big\{p(\bx_{\ell i}|\btheta_{2})\big\} -\log \big\{p_\ell(\bD_\ell)\big\}.
\end{align*}
Previously, and in the formulas above, we see that the log posterior density is decomposed into terms that depend on $\btheta_1$ or $\btheta_2$, but never on both. That means that the BFI estimator for $\btheta_1$ is not affected by the estimator $\btheta_2$ and vice versa. Therefore, in this setting, the BFI estimator  $\widehat{\btheta}_2$ can be expressed in terms of the local MAP estimators $\widehat{\btheta}_{2,\ell}$ and $\widehat{\bA}_{2,\ell}$ as in (\ref{eq:recover_theta2}). In the remainder of the derivation we focus on $\btheta_1$ only and leave out the terms with $\btheta_2$ from the expressions. 

Like in the homogeneous setting, the log posterior density in the full data set can be written in terms of the local log posterior densities:
\begin{align}
\hspace*{-3mm}
\log \big\{p(\btheta|\bD)\big\} \;=\; \sum_{\ell=1}^L \log \big\{p_\ell(\btheta_{1a},\btheta_{1b,\ell}|\bD_\ell)\big\}+\log\Bigg\{\frac{p(\btheta_{1a})}{\prod_{\ell=1}^L p_\ell(\btheta_{1a})}\Bigg\} \;+ \; \log\Bigg\{\frac{\prod_{\ell=1}^L p(\btheta_{1b,\ell})}{\prod_{\ell=1}^L p_\ell(\btheta_{1b,\ell})}\Bigg\}+B
\label{eq:linkonlyB}
\end{align}
with $B$ a term that depends on the data and on $\btheta_2$, but is not a function of $\btheta_1$.

Let $\widehat{\btheta}_{1a,\ell}$ and $\widehat{\btheta}_{1b,\ell}$ be the MAP estimators of $\btheta_{1a}$ and $\btheta_{1b,\ell}$ based on the data set $\bD_\ell$. Moreover, let $\widehat{\bA}_{1a,\ell}$ and $\widehat{\bA}_{1b,\ell}$ be minus the second derivative of $\log \{p_\ell(\btheta|\bD_\ell)\}$ with respect to $\btheta_{1a}$ and $\btheta_{1b,\ell}$ respectively, and let $\widehat{\bA}_{1ab,\ell}$ be minus the second derivative with respect to both $\btheta_{1a}$ and $\btheta_{1b,\ell}$ all evaluated at $\widehat{\btheta}_{1,\ell} =(\widehat{\btheta}_{1a,\ell},\widehat{\btheta}_{1b,\ell})$. %The corresponding estimators based on the local data set $\bD_\ell$ are denoted with a ``hat'' instead of a ``tilde'': $\widehat{\btheta}_{1b,\ell}$, $\widehat{\bA}_{1b,\ell}$ and $\widehat{\bA}_{1ab,\ell}$.

The Taylor expansions up to the quadratic term of $\log \{p_\ell(\btheta_{1a},\btheta_{1b,\ell}|\bD_\ell)\}$ around  $\widehat{\btheta}_{1,\ell}$ is given by
\begin{align*}
%\log \big\{p\big(\btheta|\bD\big)\big\} &\approx\; \log \big\{p\big(\widehat{\btheta}|\bD\big)\big\}  -\tfrac{1}{2}\big(\btheta_{1a}-\widehat{\btheta}_{1a}\big)^t \widehat{\bA}_{1a}\big(\btheta_{1a}-\widehat{\btheta}_{1a}\big) \nonumber\\
%& \qquad -\;\tfrac{1}{2}\sum_{\ell=1}^L \big(\btheta_{1b,\ell}-\widetilde{\btheta}_{1b,\ell}\big)^t \widetilde{\bA}_{1b,\ell}\big(\btheta_{1b,\ell}-\widetilde{\btheta}_{1b,\ell}\big)-\big(\btheta_{1a}-\widehat{\btheta}_{1a}\big)^t\sum_{\ell=1}^L \widetilde{\bA}_{1ab,\ell}\big(\btheta_{1b,\ell}-\widetilde{\btheta}_{1b,\ell}\big),\\
\log \big\{p_\ell\big(\btheta_{1a},\btheta_{1b,\ell}|\bD_\ell\big)\big\} &=\; \log \big\{p_\ell\big(\widehat{\btheta}_{1a,\ell},\widehat{\btheta}_{1b,\ell}|\bD_\ell\big)\big\} -\tfrac{1}{2}\big(\btheta_{1a}-\widehat{\btheta}_{1a, \ell}\big)^t \widehat{\bA}_{1a, \ell}\big(\btheta_{1a}-\widehat{\btheta}_{1a, \ell}\big) \nonumber\\[5pt]
&\qquad -\;\tfrac{1}{2} \big(\btheta_{1b,\ell}-\widehat{\btheta}_{1b,\ell}\big)^t \widehat{\bA}_{1b,\ell}\big(\btheta_{1b,\ell}-\widehat{\btheta}_{1b,\ell}\big)-\big(\btheta_{1a}-\widehat{\btheta}_{1a, \ell}\big)^t \widehat{\bA}_{1ab,\ell}\big(\btheta_{1b,\ell}-\widehat{\btheta}_{1b,\ell}\big) \;+\; O_p\big(\|\widehat{\btheta}_{1,\ell} - \btheta_1\|^3\big).
\end{align*}
Next, we insert the Taylor expressions into (\ref{eq:linkonlyB}). For the combined data $\bD$ we assume a Gaussian prior with mean zero and inverse covariance matrix $\bLambda_{1a}$ for $\btheta_{1a}$, and a zero mean Gaussian prior with inverse covariance matrix $\bLambda_{1b\ell}$ for $\btheta_{1b,\ell}, \ell=1,\ldots,L$. For center $\ell$, also zero mean Gaussian priors are chosen, but with inverse covariance matrices $\bLambda_{1a,\ell}$ and $\bLambda_{1b,\ell}$. The dimension of $\bLambda_{1b\ell}$ and $\bLambda_{1b,\ell}$ depends on the number of parameters that may vary across the centers. If only the intercepts vary, the matrices are scalars. 
After inserting these densities in expression (\ref{eq:linkonlyB}) as well, we obtain
\begin{align}
\log \big\{p(\btheta|\bD)\big\} &=\; 
 -\tfrac{1}{2}\sum_{\ell=1}^L 
 \big(\btheta_{1a}-\widehat{\btheta}_{1a,\ell}\big)^t \widehat{\bA}_{1a,\ell}\big(\btheta_{1a}-\widehat{\btheta}_{1a,\ell}\big) 
 -\tfrac{1}{2}\btheta_{1a}^t\Big(\bLambda_{1a}-\sum_{\ell=1}^L\bLambda_{1a,\ell}\Big)\btheta_{1a} \;-\tfrac{1}{2}\sum_{\ell=1}^L 
  \big(\btheta_{1b,\ell}-\widehat{\btheta}_{1b,\ell}\big)^t \widehat{\bA}_{1b,\ell}\big(\btheta_{1b,\ell}-\widehat{\btheta}_{1b,\ell}\big)
\nonumber\\
 &~~~~  \; -\tfrac{1}{2} \sum_{\ell=1}^L \btheta_{1b,\ell}^t\big(\bLambda_{1b\ell}-\bLambda_{1b,\ell}\big)\btheta_{1b,\ell}  \;-\;  \sum_{\ell=1}^L 
 \big(\btheta_{1a}-\widehat{\btheta}_{1a,\ell}\big)^t \widehat{\bA}_{1ab,\ell}\big(\btheta_{1b,\ell}-\widehat{\btheta}_{1b,\ell}\big)
 \;+\; B' \;+\; O_p\Big(\sum_{\ell=1}^L \|\widehat{\btheta}_{1,\ell}-\btheta_1\|^3\Big)
\nonumber\\
&=:\; \Omega_{\BFI}(\btheta_1) +  O_p\Big(\sum_{\ell=1}^L \|\widehat{\btheta}_{1,\ell}-\btheta_1\|^3\Big)
\label{eq:equation}
\end{align}
with $B'$ representing a term that does not depend on $\btheta_1$. The function $\btheta_1\rightarrow \Omega_{\BFI}(\btheta_1)$ is a quadratic function of $\btheta_1$. Maximization of this function with respect to $\btheta_1$ by setting its derivative equal to zero, yields the BFI estimators
\begin{align}
\widehat{\btheta}_{1a,\BFI}&:=\; \Big( \widehat{\bA}_{1a,\BFI}
 -\sum_{\ell=1}^L \widehat{\bA}_{1ab,\ell}(\widehat{\bA}_{1b,\ell,\BFI})^{-1}(\widehat{\bA}_{1ab,\ell,\BFI})^t
 \Big)^{\!-1}\times \nonumber\\ 
& ~~~~~~~~ \sum_{\ell=1}^L \Big[
  \Big(
 \widehat{\bA}_{1a,\ell}
 -
 \widehat{\bA}_{1ab,\ell}(\widehat{\bA}_{1b,\ell,\BFI})^{-1}(\widehat{\bA}_{1ab,\ell})^t\Big)\widehat{\btheta}_{1a,\ell} 
  + 
  \widehat{\bA}_{1ab,\ell}\Big({\bf 1}
  -
(\widehat{\bA}_{1b,\ell,\BFI})^{-1}\widehat{\bA}_{1b,\ell}\Big)\widehat{\btheta}_{1b,\ell}
\Big]
\label{eq:theta1ahat}
\end{align}
with ${\bf 1}$ the unit matrix and
the matrices $\widehat{\bA}_{1a,\BFI}$ and $\widehat{\bA}_{1b,\ell,\BFI}$ as given in (\ref{eq:with_nuisance_B}) below and
\begin{align}
\widehat{\btheta}_{1b,\ell,\BFI} 
\; :=\;\big(\widehat{\bA}_{1b,\ell,\BFI}\big)^{-1}\Big[\widehat{\bA}_{1b,\ell}\widehat{\btheta}_{1b,\ell}
 + \big(\widehat{\bA}_{1ab,\ell}\big)^t\big(\widehat{\btheta}_{1a,\ell} 
 -
 \widehat{\btheta}_{1a,\BFI}\big)\Big] 
\label{eq:theta1bhat}
 \end{align}
with
\begin{align}
\label{eq:with_nuisance_B}
\widehat{\bA}_{1a,\BFI}:=\sum_{\ell=1}^L \widehat{\bA}_{1a,\ell}+\bLambda_{1a}-\sum_{\ell=1}^L\bLambda_{1a,\ell},
~~~~~~~
\widehat{\bA}_{1b,\ell, \BFI}:=\widehat{\bA}_{1b,\ell}+\bLambda_{1b\ell}-\bLambda_{1b,\ell},
~~~~~~~
\widehat{\bA}_{1ab,\ell, \BFI}:=\widehat{\bA}_{1ab,\ell},
\end{align}
where $\widehat{\bA}_{1a,\BFI}, \widehat{\bA}_{1b,\ell, \BFI}$ and $\widehat{\bA}_{1ab,\ell, \BFI}$ equal minus the second derivatives of $\Omega_{\BFI}$ with respect to $\btheta_{1a}, \btheta_{1b}$ and the mix $\btheta_{1a}$ and $\btheta_{1b}$.

\subsection*{Appendix II.C: Heterogeneity across centers due to clustering}
\label{subsec: Appendix C}

 { In this appendix we consider the situation in which the centers are clustered by, for example, due to location or type (academic / non-academic medical center). Another example is the covariate hospital size in the nurse-data set, where the clusters are: small, medium, large. Within a hospital/center, all nurses are in the same cluster and have the same covariate value, so that the covariate can not be included in the local regression model as it is collinear with the intercept. } 

 { In the calculations of the BFI estimators, we assume that the covariates are independent between the individuals within and across the centers. We, moreover, assume that the outcome variables given the covariates and the cluster level are independent.}
Suppose that the vector of model parameters in center $\ell$ is equal to $\btheta_\ell = (\btheta_{1a},\theta_{1b,\ell},\btheta_2)$, where, as before, $\btheta_2$ is the parameter vector that specifies the distribution of the covariates. The parameter $\btheta_{1a}$ is the vector of regression parameters which are assumed to be equal in all centers, but excluding the intercept which may vary across the centers. The parameter $\theta_{1b,\ell} \in \{\theta_{1b1},\ldots,\theta_{1bK}\}$ for $\ell=1,\ldots,L$ and with $K\leq L$ is the intercept of the model in center $\ell$. So, $\theta_{1b,\ell}$ (with a comma in the subscript) is the parameter in center $\ell$, whereas $\theta_{1bk}$ (without a comma in the subscript) is the parameter for the $k^{th}$ category of the center-specific covariate. 
If $K=L$,  $\theta_{1b,\ell}\neq \theta_{1b,\ell'}$ for $\ell \neq \ell'$ and we are in the situation of Appendix II.B, where every center has its own specific intercept value. If $K<L$, there are centers $\ell$ and $\ell'$ with $\ell\neq \ell'$ with $\theta_{1b,\ell}= \theta_{1b,\ell'}$. In the example, the covariate ``hospital size'' has three levels (small, medium, large). That means that $K=3$ and the three parameters represent the three intercepts for the three classes of centers. 
The parameter vector in the (fictive) combined data set $\bD$ is defined as $\btheta=(\btheta_{1a}, \theta_{1b1}, \ldots, \theta_{1bK}, \btheta_2)$. 

For simplicity of notation we assume (again) that $\btheta_{1a}, \theta_{1b}$ and $\btheta_2$ are independent:  $p(\btheta)=p(\btheta_{1a})p(\btheta_{2})\prod_{k=1}^K p(\theta_{1bk})$ for the combined data set, and locally $p_\ell(\btheta_{1a},\theta_{1b,\ell},\btheta_{2})=p_\ell(\btheta_{1a})p_\ell(\theta_{1b,\ell})p_\ell(\btheta_{2})$ in data subset $\ell$.
For the combined data $\bD$ we assume a Gaussian prior with mean zero and inverse covariance matrices $\bLambda_{1a}$ for $\btheta_{1a}$, and a zero mean Gaussian prior with inverse variance $\bLambda_{1bk}$ for $\btheta_{1bk}, k=1,\ldots,K$. Also for center $\ell$ zero mean Gaussian priors are chosen, but with inverse covariance matrix  $\bLambda_{1a,\ell}$ and inverse variance $\bLambda_{1b,\ell}$. Similar notation and calculations as in Appendix II.B lead to the equation below, instead of the equation (\ref{eq:equation}): 
\begin{align*}
\lefteqn{ \log \big\{p(\btheta|\bD)\big\}}\\ &=\;
 -\tfrac{1}{2}\sum_{\ell=1}^L 
 \big(\btheta_{1a}-\widehat{\btheta}_{1a,\ell}\big)^t \widehat{\bA}_{1a,\ell}\big(\btheta_{1a}-\widehat{\btheta}_{1a,\ell}\big) 
\;-\;\tfrac{1}{2}\btheta_{1a}^t\Big(\bLambda_{1a}-\sum_{\ell=1}^L\bLambda_{1a,\ell}\Big)\btheta_{1a} \;-\;\tfrac{1}{2}\sum_{\ell=1}^L 
  \big(\theta_{1b}-\widehat{\theta}_{1b,\ell}\big) \widehat{A}_{1b,\ell}\big(\theta_{1b,\ell}-\widehat{\theta}_{1b,\ell}\big) \nonumber
\\
& \hspace{5mm} 
   \;-\;\tfrac{1}{2} \sum_{k=1}^K \theta_{1bk}\Lambda_{1bk}\theta_{1bk} \;+\;\tfrac{1}{2} \sum_{\ell=1}^L \theta_{1b,\ell}\Lambda_{1b,\ell}\theta_{1b,\ell}  \;-\;  \sum_{\ell=1}^L 
 \big(\btheta_{1a}-\widehat{\btheta}_{1a,\ell}\big)^t \widehat{\bA}_{1ab,\ell}\big(\btheta_{1b,\ell}-\widehat{\btheta}_{1b,\ell}\big)
 \;+\; B' \;+\; O_p\Big(\sum_{\ell=1}^L \|\widehat{\btheta}_{1,\ell}-\btheta_1\|^3\Big)\\
&=:\; \Omega_{\BFI}(\btheta_1) \;+\; O_p\Big(\sum_{\ell=1}^L \|\widehat{\btheta}_{1,\ell}-\btheta_1\|^3\Big)  
\end{align*}
with $B'$ representing a term that does not depend on $\btheta_1$. Let $z_\ell$ denote the category of center $\ell$ for the center-specific covariate. So $z_\ell \in \{1,\ldots,K\}$. 
Differentiating $\Omega_{\BFI}(\btheta_1)$ with respect to  $\btheta_{1a}$ and $\theta_{1b}$ and setting the derivatives equal to zero, yields the BFI estimators: 
\begin{align*}
\widehat{\btheta}_{1a,\BFI}
 &:=\; \Big(\widehat{\bA}_{1a,\BFI}-\sum_{k=1}^K\widehat{\bA}_{1abk,\BFI}\big(\widehat{A}_{1bk,\BFI}\big)^{-1}\;(\widehat{\bA}_{1abk,\BFI})^t\Big)^{-1} \times\\ 
& \Bigg(
 \sum_{\ell=1}^L 
\widehat{\bA}_{1a,\ell}\widehat{\btheta}_{1a,\ell} 
+\sum_{\ell=1}^L 
  \widehat{\bA}_{1ab,\ell}\widehat{\theta}_{1b,\ell}-\sum_{k=1}^K \widehat{\bA}_{1abk,\BFI}\big(\widehat{A}_{1bk,\BFI}\big)^{-1}\Big[\sum_{\ell=1: z_\ell=k}^L\widehat{A}_{1b,\ell}\widehat{\theta}_{1b,\ell}
 +\sum_{\ell=1:z_\ell=k}^L(\widehat{\bA}_{1ab,\ell})^t\widehat{\btheta}_{1a,\ell}  \Big]\Bigg)
 \end{align*}
with $\widehat{\bA}_{1a,\BFI}$,  $\widehat{A}_{1bk,\BFI}$ and  $\widehat{\bA}_{1abk,\BFI}$ as given in (\ref{eq:with_nuisance_1}) below and the estimator $\widehat{\theta}_{1bk,\BFI}, k=1,\ldots,K$ is given by  
\begin{eqnarray*}
\widehat{\theta}_{1bk,\BFI} 
\;:=\;\big(\widehat{A}_{1bk,\BFI}\big)^{-1}\Big[\sum_{\ell=1: z_\ell=k}^L\widehat{A}_{1b,\ell}\widehat{\theta}_{1b,\ell}
 \;+\;\sum_{\ell=1: z_\ell=k}^L(\widehat{\bA}_{1ab,\ell})^t\widehat{\btheta}_{1a,\ell} \;-\;(\widehat{\bA}_{1abk,\BFI})^t\widehat{\btheta}_{1a,\BFI} \Big]. 
 \end{eqnarray*}
with minus the second derivatives of $\Omega_{\BFI}$ equal to
\begin{align}
\widehat{\bA}_{1a,\BFI} &:=\; \sum_{\ell=1}^L \widehat{\bA}_{1a,\ell}\;+\;\bLambda_{1a}\;-\;\sum_{\ell=1}^L\bLambda_{1a,\ell}, \nonumber\\
\widehat{A}_{1bk,\BFI} &:=\;\sum_{\ell=1: z_\ell=k}^L \widehat{A}_{1b,\ell}\;+\;\Lambda_{1bk}\;-\;\sum_{\ell=1: z_\ell=k}^L\Lambda_{1b,\ell}, \label{eq:with_nuisance_1}\\
\widehat{\bA}_{1abk,\BFI} &:=\; \sum_{\ell=1: z_\ell=k}^L\widehat{\bA}_{1ab,\ell}, \nonumber
\end{align}
and 0 for the remaining terms.

\section*{Appendix III: asymptotic theory of the BFI and WAV estimators}
In this appendix we compute the asymptotic distribution of the BFI and WAV estimators under the assumption of homogeneity and heterogeneity. In the calculations we assume that the number of clusters $L$ is fixed, but the sample sizes within the clusters, $n_1,\ldots,n_L$, increase to infinity such that, for $n=n_1+\ldots+n_L$, the fraction $n_\ell/n \rightarrow w_\ell$, with $0\leq w_\ell \leq 1, \ell=1,\ldots,L$. In all cases we assume no model-misspecification and the independence assumptions stated in Appendix II. This Appendix consists of
\begin{enumerate}
\item[] Appendix III.A: Asymptotic distribution of the MAP estimator based on the combined data. 
\item[] Appendix III.B: Asymptotic distribution of the BFI and WAV estimators in a homogeneous setting.
\item[] Appendix III.C: Asymptotic distribution of the BFI and WAV estimators in a heterogeneous setting. 
\end{enumerate}

\subsection*{Appendix III.A: Asymptotic distribution of the MAP estimator based on the combined data}
In this section we study the asymptotic distribution of the MAP estimator $\widehat{\btheta}_1=(\widehat{\btheta}_{1a},\widehat{\btheta}_{1b})$ for $\btheta_1=(\btheta_{1a},\btheta_{1b})$ in the combined data set. The asymptotic distribution for the MAP estimator for $\btheta_2$ can be derived similarly. 

From literature (Bernstein-Von Mises Theorem \cite{vdVaart}) it is known that the MAP estimator is asymptotically Gaussian:
\begin{align*}
\sqrt{n} \begin{pmatrix} \begin{pmatrix}
\widehat{\btheta}_{1a}\\
\widehat{\btheta}_{1b}
\end{pmatrix} -
\begin{pmatrix}
\btheta_{1a}\\
\btheta_{1b}
\end{pmatrix}
\end{pmatrix} \;\leadsto \; {\mathcal N}\big({\bf 0}, J_{1}^{-1} \big),
\end{align*}
where "$\leadsto$" means convergence in distribution for the sample size to infinity. Further, ${\bf 0}$ is a vector of zeroes, and $J_{1}^{-1}$ is the inverse Fisher information matrix for the combined populations from all centers. The Fisher information matrix and its inverse have the form
\begin{align*}
J_{1} &=\; 
\begin{pmatrix}
J_{1a} & J_{1ab} \\
(J_{1ab})^t & J_{1b}
\end{pmatrix}\\[10pt]
(J_{1})^{-1} &=\; 
\begin{pmatrix}
\big(J_{1a}-J_{1ab} (J_{1b})^{-1} (J_{1ab})^t\big)^{-1} & ~~~ -\big(J_{1a}-J_{1ab} (J_{1b})^{-1} (J_{1ab})^t\big)^{-1}J_{1ab} (J_{1b})^{-1} \\
-(J_{1b})^{-1} (J_{1ab})^t \big(J_{1a}-J_{1ab} (J_{1b})^{-1} (J_{1ab})^t\big)^{-1} & ~~~ (J_{1b})^{-1} + (J_{1b})^{-1} (J_{1ab})^t \big(J_{1a}-J_{1ab} (J_{1b})^{-1} (J_{1ab})^t\big)^{-1} J_{1ab} (J_{1b})^{-1}
\end{pmatrix}
\end{align*}
where $J_{1a}$ and $J_{1b}$ equal the expectation of the second derivatives of $-\log \{p(\bD|\btheta)\}$ with respect to $\btheta_{1a}$ and $\btheta_{1b}$, evaluated at the true value of $\btheta_1$. The matrix $J_{1ab}$ equals the expectation of the derivative of $-\log \{p(\bD|\btheta)\}$ with respect to $\btheta_{1a}$ and $\btheta_{1b}$, evaluated at the true value of $\btheta_1$.  

In the following we rewrite the four submatrices of $(J_1)^{-1}$ in terms of the Fisher information matrices in the local centers. In the next appendices it will be proven that the asymptotic covariance matrices of the BFI-estimators equal these expressions and the BFI estimators are therefore asymptotically efficient.

The data from the different centers are independent. Therefore, the Fisher information matrix can be written as a weighted sum of the Fisher information matrices in the different centers. By the law of large numbers 
\begin{align*}
J_1 = \lim_{n_1,\ldots,n_L\rightarrow \infty}  \frac{\partial^2}{\partial \btheta_1^2} \Big(- \; \frac{1}{n}\log \big\{p(\bD | \btheta_1)\big\}\Big) &= \lim_{n_1,\ldots,n_L\rightarrow \infty}  \frac{\partial^2}{\partial \btheta_1^2} \Big(- \; \sum_{\ell=1}^L \frac{1}{n}\log \big\{p(\bD_\ell | \btheta_1)\big\}\Big)\\[5pt]
&= \lim_{n_1,\ldots,n_L\rightarrow \infty} \; \sum_{\ell=1}^L \frac{n_\ell}{n} \frac{\partial^2}{\partial \btheta_1^2} \Big(-\frac{1}{n_\ell}\log \big\{p(\bD_\ell | \btheta_1)\big\}\Big) = \sum_{\ell=1}^L w_\ell J_{1,\ell}  
\end{align*}
with $J_{1,\ell}$ the Fisher information matrix for $\btheta_1$ in center $\ell$ and $n_\ell/n \rightarrow w_\ell$ if $n_\ell, n \rightarrow \infty$. So $J_1 = \sum_{\ell=1}^L w_\ell J_{1,\ell}$.

In the homogeneous setting all parameters are included in $\btheta_{1a}$ (there is no parameter $\btheta_{1b}$). Then, for $I_{1,\ell}$ the Fisher information matrix for $\btheta_{1a}$ in center $\ell$, it follows that $J_{1,\ell}=I_{1,\ell}=I_1, \ell=1,2,\ldots,L$ and 
\begin{align}
J_1 \;=\; \sum_{\ell=1}^L w_\ell J_{1,\ell} \;=\; \sum_{\ell=1}^L w_\ell I_{1,\ell} \;=\; I_1, ~~~~~~~~~  (J_1)^{-1} \;=\; \Big(\sum_{\ell=1}^L w_\ell I_{1,\ell}\Big)^{-1} \;=\; (I_1)^{-1}.   
\label{eq: J1inv}
\end{align}

The heterogeneous setting is more complex. Suppose the parameter $\btheta_{1a}$ is assumed to be same across all centers, but $\btheta_{1b}=(\btheta_{1b,1},\ldots,\btheta_{1b,L})$ is a vector with center-specific parameters (the index refers to the center). The log likelihood function for center $\ell$ is a function of $\btheta_{1b,\ell}$, but not of $\btheta_{1b,k}$ with $k\neq \ell$. Therefore, for the Fisher information matrix $J_{1,\ell}$ for $(\btheta_{1a},\btheta_{1b})=(\btheta_{1a},\btheta_{1b,1},\ldots,\btheta_{1b,L})$ in center $\ell$, the columns and rows that are related to  $\btheta_{1b,k}, k\neq \ell$ contain zeroes only. The matrix $I_{1,\ell}$ is the Fisher information matrix for $(\btheta_{1a},\btheta_{1b,\ell})$ in center $\ell$ (so not of $(\btheta_{1a},\btheta_{1b})$ like $J_{1,\ell}$), with the blocks $I_{1a,\ell}, I_{1b,\ell}$ and $I_{1ab,\ell}$, defined in a similar way as in $J_1$. Since $\btheta_{1a}$ is the same across the centers,  $J_{1a,\ell}=I_{1a,\ell}$. However, $J_{1b,\ell}\neq I_{1b,\ell}$, since $J_{1b,\ell}$ is the Fisher information matrix for $\btheta_{1b}=(\btheta_{1b,1}, \ldots, \btheta_{1b,L})$ in center $\ell$, whereas $I_{1b,\ell}$ is the Fisher information matrix for $\btheta_{1b,\ell}$ in center $\ell$; the dimensions of the matrices are different. 

Since the parameter $\btheta_{1b}$ is a vector with center-specific parameters, the matrix $J_{1b}$ has a block diagonal matrix with center-specific blocks. Because of this form, it follows that
\begin{align*}
J_{1ab} (J_{1b})^{-1} (J_{1ab})^t = \sum_{\ell=1}^L w_{\ell} I_{1ab,\ell} (I_{1b,\ell})^{-1} (I_{1ab,\ell})^t.
\end{align*}
Since $J_{1a}=\sum_{\ell=1}^L w_\ell J_{1a,\ell}=\sum_{\ell=1}^L w_\ell I_{1a,\ell}$ (the parameter vector $\btheta_{1a}$ is shared across all centers), 
\begin{align*}
J_{1a}-J_{1ab} (J_{1b})^{-1} (J_{1ab})^t = \sum_{\ell=1}^L w_\ell \big(I_{1a,\ell}-I_{1ab,\ell} (I_{1b,\ell})^{-1} (I_{1ab,\ell})^t \big)
\end{align*}
and the asymptotic covariance matrix for $\widehat{\btheta}_{1a}$ is equal to
\begin{align}
\Big(J_{1a}-J_{1ab} (J_{1b})^{-1} (J_{1ab})^t\Big)^{-1} = \Big(\sum_{\ell=1}^L w_\ell \big(I_{1a,\ell}-I_{1ab,\ell} (I_{1b,\ell})^{-1} (I_{1ab,\ell})^t \big)\Big)^{-1}.
\label{Eq: inv cov}
\end{align}
%If the matrices $I_{1,\ell}=I_1, \ell=1,\ldots,L$, 
%\begin{align*}
%J_{1a}-J_{1ab} (J_{1b})^{-1} (J_{1ab})^t = I_{1a}-I_{1ab} (I_{1b})^{-1} (I_{1ab})^t
%\end{align*}
%and the asymptotic covariance matrix for $\widehat{\btheta}_{1a}$ simplifies to $(I_{1a}-I_{1ab} (I_{1b})^{-1} (I_{1ab})^t)^{-1}$.

The asymptotic covariance matrix for the MAP estimator $\widehat{\btheta}_{1b}$ equals:
\begin{align}
(J_{1b})^{-1} + (J_{1b})^{-1} (J_{1ab})^t \big(J_{1a}-J_{1ab} (J_{1b})^{-1} (J_{1ab})^t\big)^{-1} J_{1ab} (J_{1b})^{-1}.
\label{eq: ellblock}
\end{align}
For parameter $\btheta_{1b,\ell}$ the asymptotic covariance matrix equals the $\ell^{th}$ diagonal block of this matrix. The corresponding block of the matrix $J_{1b}$ equals $w_\ell I_{1b,\ell}$. 
By the structure of $J_{1b}$ and the equation (\ref{Eq: inv cov}) it follows that $\ell^{th}$ diagonal block of the matrix in (\ref{eq: ellblock}) is given by
\begin{align}
\label{eq: asympt 1b}
 \big(w_\ell I_{1b,\ell}\big)^{-1}+\big(I_{1b,\ell}\big)^{-1} \big(I_{1ab,\ell}\big)^{t} \Big(\sum_{k=1}^L w_k \big(I_{1a,k}-I_{1ab,k}(I_{1b,k})^{-1}(I_{1ab,k}\big))^t\Big)^{-1} I_{1ab,\ell} \big(I_{1b,\ell}\big)^{-1}.
\end{align}

\subsection*{Appendix III.B: Asymptotic distribution in homogeneous setting}

\subsubsection*{Asymptotic distribution BFI estimator}
In Appendix II.A we have derived an expression for the BFI estimator in the homogeneous setting. In this setting, all parameters are included in the vector $\btheta_{1a}$; there is no vector $\btheta_{1b}$. The BFI estimator $\widehat{\btheta}_{1,\BFI}$ is defined as  
\begin{align*}
\widehat{\btheta}_{1,\BFI} =  \big(\widehat{\bA}_{1,\BFI}\big)^{-1}\sum_{\ell=1}^L \widehat{\bA}_{1,\ell}\widehat{\btheta}_{1,\ell} ~~~~~~~~~~~\mbox{with} ~~~~~~~~~~\widehat{\bA}_{1,\BFI} = \sum_{\ell=1}^L \widehat{\bA}_{1,\ell}+\bLambda_1-\sum_{\ell=1}^L \bLambda_{1,\ell}.
\end{align*}
Below, we derive the asymptotic distribution of $\sqrt{n}\big(\widehat{\btheta}_{1,\BFI} - \btheta_1\big)$: 
\begin{align*}
\sqrt{n}\big(\widehat{\btheta}_{1,\BFI} - \btheta_1\big) &= \sqrt{n} \Big\{ \Big(\sum_{\ell=1}^L \widehat{\bA}_{1,\ell}+\bLambda_1-\sum_{\ell=1}^L \bLambda_{1,\ell}\Big)^{-1}\Big(\sum_{\ell=1}^L \widehat{\bA}_{1,\ell}\widehat{\btheta}_{1,\ell}\Big)-\btheta_1\Big\} \\[4pt] 
&= \sqrt{n} \Big\{ \Big(\sum_{\ell=1}^L \frac{1}{n}\widehat{\bA}_{1,\ell}+\frac{1}{n}\bLambda_1-\sum_{\ell=1}^L \frac{1}{n}\bLambda_{1,\ell}\Big)^{-1}\Big(\sum_{\ell=1}^L \frac{1}{n}\widehat{\bA}_{1,\ell}\widehat{\btheta}_{1,\ell}\Big)-\btheta_1\Big\} 
\end{align*}
Because the term
\begin{align*}
\Big(\sum_{\ell=1}^L \frac{1}{n}\widehat{\bA}_{1,\ell}+\frac{1}{n}\bLambda_1-\sum_{\ell=1}^L \frac{1}{n}\bLambda_{1,\ell}\Big)^{-1} =  \Big(\sum_{\ell=1}^L \frac{1}{n}\widehat{\bA}_{1,\ell}\Big)^{-1} + O_p\Big(\frac{1}{n}\Big),     
\end{align*}
it holds that
\begin{align*}
\sqrt{n}\big(\widehat{\btheta}_{1,\BFI} - \btheta_1\big) &= \sqrt{n} \Big\{ \Big(\sum_{\ell=1}^L \frac{1}{n}\widehat{\bA}_{1,\ell}\Big)^{-1}\Big(\sum_{\ell=1}^L \frac{1}{n}\widehat{\bA}_{1,\ell}\widehat{\btheta}_{1,\ell}\Big)-\btheta_1\Big\} + O_p\Big(\frac{1}{\sqrt{n}}\Big)\\[4pt]
&= \Big(\sum_{\ell=1}^L \frac{1}{n}\widehat{\bA}_{1,\ell}\Big)^{-1}\Big(\sum_{\ell=1}^L \frac{1}{n}\widehat{\bA}_{1,\ell} \;\sqrt{n}\big(\widehat{\btheta}_{1,\ell}-\btheta_1\big)\Big) + O_p\Big(\frac{1}{\sqrt{n}}\Big)\\[4pt]
&= \Big(\sum_{\ell=1}^L \frac{n_\ell}{n} \frac{1}{n_\ell}\widehat{\bA}_{1,\ell}\Big)^{-1}\Big(\sum_{\ell=1}^L \frac{\sqrt{n_\ell}}{\sqrt{n}} \; \frac{1}{n_\ell}\widehat{\bA}_{1,\ell} \;\sqrt{n_\ell}\big(\widehat{\btheta}_{1,\ell}-\btheta_1\big)\Big) + O_p\Big(\frac{1}{\sqrt{n}}\Big).
\end{align*}

Asymptotically, the MAP estimator and the maximum likelihood estimator are equivalent. It follows that the MAP estimator in center $\ell$, $\widehat{\btheta}_{1,\ell}$, is asymptotically normal\cite{vdVaart, Bijma}:  $\sqrt{n_\ell}(\widehat{\btheta}_{1,\ell}-\btheta_{1,\ell}) \leadsto {\mathcal N}(0,(I_{1,\ell})^{-1})$ for $I_{1,\ell}$ the Fisher information matrix in center $\ell$. Remember that $\widehat{\bA}_{1,\ell}$ is defined as the second derivative of $-\log \{p(\bD_\ell|\btheta_1)\}$ evaluated at $\widehat{\btheta}_{1,\ell}$. If this second derivative is sufficiently smooth near $\btheta_{1,\ell}$, it follow by the law of large numbers, that  $n_\ell^{-1}\widehat{\bA}_{1,\ell}$ converges in probability to $I_{1,\ell}$.\cite{vdVaart, Bijma} 
By Slutsky's lemma, it follows that, for every center $\ell$
\begin{align*}
\frac{1}{n_\ell}\widehat{\bA}_{1,\ell} \; \sqrt{n_\ell}\big(\widehat{\btheta}_{1,\ell}-\btheta_1\big) \;=\;  I_{1,\ell} \; \sqrt{n_\ell}\big(\widehat{\btheta}_{1,\ell}-\btheta_1\big) + o_P(1) \;\leadsto\; {\mathcal N}\big(0, I_{1,\ell}\big).
\end{align*}
Since the data across the $L$ centers are assumed to be independent, it follows that
\begin{align*}
\sum_{\ell=1}^L \frac{\sqrt{n_\ell}}{\sqrt{n}} \; \frac{1}{n_\ell}\widehat{\bA}_{1,\ell} \; \sqrt{n_\ell}\big(\widehat{\btheta}_{1,\ell}-\btheta_1\big) \;=\; \sum_{\ell=1}^L \sqrt{w_\ell} \; I_{1,\ell} \; \sqrt{n_\ell}\big(\widehat{\btheta}_{1,\ell}-\btheta_1\big) + o_P(1) \;\leadsto\; {\mathcal N}\Big(0,\sum_{\ell=1}^L w_\ell I_{1,\ell}\Big).
\end{align*}
Further, the term 
\begin{align*}
\Big(\sum_{\ell=1}^L \frac{n_\ell}{n} \frac{1}{n_\ell}\widehat{\bA}_{1,\ell}\Big)^{-1} \;=\; \Big(\sum_{\ell=1}^L w_\ell I_{1,\ell}\Big)^{-1}  \;+\; o_p(1).    
\end{align*}
Combining the results, yields
\begin{align*}
\sqrt{n}\big(\widehat{\btheta}_{1,\BFI} - \btheta_1\big) \leadsto {\mathcal N}\Big(0,\Big(\sum_{\ell=1}^L w_\ell I_{1,\ell}\Big)^{-1}\Big).
\end{align*}
The asymptotic covariance matrix equals $J_1^{-1}$ as defined in (\ref{eq: J1inv}), which equals the asymptotic covariance matrix of the MAP estimator based on the combined data. The BFI estimator is asymptotically efficient; no information is lost if the data from the centers can not be combined. Under homogeneity the matrices $I_{1,\ell}=I_1=J_1, \ell=1,\ldots,L$, and because $\sum_{\ell=1}^L w_\ell =1$, 
\begin{align*}
\sqrt{n}\big(\widehat{\btheta}_{1,\BFI} - \btheta_1\big) \leadsto {\mathcal N}\Big(0,\big(I_{1}\big)^{-1}\Big).
\end{align*}

Further, we have seen that the BFI estimator  
\begin{align*}
\frac{1}{n}\widehat{\bA}_{1,\BFI} \;=\; \frac{1}{n}\sum_{\ell=1}^L \widehat{\bA}_{1,\ell}+\frac{1}{n}\bLambda_1-\frac{1}{n}\sum_{\ell=1}^L \bLambda_{1,\ell} \;=\; \sum_{\ell=1}^L \frac{n_\ell}{n}\; \frac{1}{n_\ell}\widehat{\bA}_{1,\ell}\;+\;O_p\Big(\frac{1}{n}\Big)
\;=\; \sum_{\ell=1}^L w_\ell I_{1,\ell} \;+\;o_p(1),
\end{align*} 
converges in probability to $\sum_{\ell=1}^L w_\ell I_{1,\ell}$. 

 { If the number of centers $L$ increases to infinity as well, but $L=o_p(n)$ (i.e., the number of centers is smaller in rate than the total sample size), the asymptotic results remain valid, but the derivation needs to be adjusted slightly. }

\subsubsection*{Asymptotic distribution of the WAV and Single center estimators}
Suppose the MAP estimator $\widehat{\btheta}_{1,\ell}$ in center $\ell$ is used for estimating the parameter $\btheta_1$, then we obtain
\begin{align*}
\sqrt{n}\big(\widehat{\btheta}_{1,\ell} - \btheta_1\big) \;=\; \frac{\sqrt{n}}{\sqrt{n_\ell}} \; \sqrt{n_\ell}\big(\widehat{\btheta}_{1,\ell} - \btheta_1\big) \leadsto {\mathcal N}\Big(0,\big(w_\ell I_{1,\ell}\big)^{-1}\Big),
\end{align*}
if $w_\ell > 0$.
That means that the single-center estimator is not optimal for estimating $\btheta_1$ unless $w_\ell=1$ (there is only one center). 

For the weighted average estimator
$\sum_{\ell=1}^L \frac{n_\ell}{n} \widehat{\btheta}_{1,\ell}$
for estimating $\btheta_1$, we find
\begin{align*}
\sqrt{n}\Big(\sum_{\ell=1}^L \frac{n_\ell}{n}\; \widehat{\btheta}_{1,\ell}-\btheta_1\Big) \;=\; \sum_{\ell=1}^L  \frac{n_\ell}{n}\; \sqrt{n} \big(\widehat{\btheta}_{1,\ell}-\btheta_1\big) \;=\; \sum_{\ell=1}^L   \frac{\sqrt{n_\ell}}{\sqrt{n}} \sqrt{n_\ell} \big(\widehat{\btheta}_{1,\ell}-\btheta_1\big) \leadsto  {\mathcal N}\Big(0,\sum_{\ell=1}^Lw_\ell \big( I_{1,\ell}\big)^{-1}\Big),
\end{align*}
with, in the homogeneous setting $I_{1,\ell}=I_1, \ell=1,\ldots,L$
\begin{align*}
\sum_{\ell=1}^Lw_\ell \big( I_{1,\ell}\big)^{-1}=(I_1)^{-1} = (J_1)^{-1}
\end{align*}
as defined in (\ref{eq: J1inv}). In the homogeneous setting, the weighted average estimator is asymptotically efficient as well.

\subsection*{Appendix III.C: Asymptotic distribution of the BFI estimator in heterogeneous setting}

\subsubsection*{Asymptotic distribution BFI estimator}
Let $\widehat{\bA}_{1a,\ell}$ and $\widehat{\bA}_{1b,\ell}$ be the second derivatives of $-\log \{p_\ell(\btheta_1 | \bD_\ell)\}$ with respect to $\btheta_{1a}$ and $\btheta_{1b}$, respectively, and let $\widehat{\bA}_{1ab,\ell}$ be the second derivative with respect to both $\btheta_{1a}$ and $\btheta_{1b}$, all evaluated at the MAP estimator $\widehat{\btheta}_{1,\ell}$. If these second derivatives are sufficiently smooth in the neighborhood of $\btheta_{1}$, it follows by the law of large numbers that
\begin{align*}
\frac{1}{n_\ell}\widehat{\bA}_{1a,\ell}  \rightarrow I_{1a,\ell} ~~~~~~~~ \frac{1}{n_\ell}\widehat{\bA}_{1b,\ell}  \rightarrow I_{1b,\ell} ~~~~~~~~ \frac{1}{n_\ell}\widehat{\bA}_{1ab,\ell}  \rightarrow I_{1ab,\ell},    
\end{align*}
with the matrices $I_{1a,\ell}, I_{1b,\ell}$ and $I_{1ab,\ell}$ for center $\ell$.

\bigskip

\noindent
{\bf Asymptotic behaviour of the BFI estimator for $\btheta_{1a}$ }\\
In Appendix II.B we computed that, under the assumption that $\Lambda_{1b\ell}=\Lambda_{1b,\ell}$, the BFI estimator for $\btheta_{1a}$ is equal to:
\begin{align}
\widehat{\btheta}_{1a,\BFI}:=\Big( \sum_{\ell=1}^L \Big(\widehat{\bA}_{1a,\ell}- \widehat{\bA}_{1ab,\ell}(\widehat{\bA}_{1b,\ell})^{-1}(\widehat{\bA}_{1ab,\ell})^t\Big) +\bLambda_{1a}-\sum_{\ell=1}^L\bLambda_{1a,\ell} 
 \Big)^{\!-1}  \sum_{\ell=1}^L \Big(\widehat{\bA}_{1a,\ell} -
 \widehat{\bA}_{1ab,\ell}(\widehat{\bA}_{1b,\ell})^{-1}(\widehat{\bA}_{1ab,\ell})^t\Big)\widehat{\btheta}_{1a,\ell}.
\end{align}
Define $\widehat{\Gamma}_\ell := n_\ell^{-1}\widehat{\bA}_{1a,\ell}- n_\ell^{-1}\widehat{\bA}_{1ab,\ell}(n_\ell^{-1}\widehat{\bA}_{1b,\ell})^{-1}(n_\ell^{-1}\widehat{\bA}_{1ab,\ell})^t$. 
Then,
\begin{align}
\widehat{\btheta}_{1a,\BFI}=\Big(\sum_{\ell=1}^L \frac{n_\ell}{n}\;\widehat{\Gamma}_\ell + \frac{1}{n}\;\bLambda_{1a}-\sum_{\ell=1}^L\frac{1}{n}\;\bLambda_{1a,\ell} 
 \Big)^{\!-1}  \sum_{\ell=1}^L \frac{n_\ell}{n}\;\widehat{\Gamma}_\ell \;\widehat{\btheta}_{1a,\ell}.
\end{align}
If the derivatives are sufficiently smooth, it follows by the law of large numbers, continuous mapping theorem, and Slutsky's lemma\cite{vdVaart}, that $\widehat{\Gamma}_\ell^{-1}$ converges in probability to $\big(I_{1a,\ell}-I_{1ab,\ell}(I_{1b,\ell})^{-1}(I_{1ab,\ell})^t\big)^{-1}$, which equals the left-upper block of the inverse of the Fisher information matrix $I_{1,\ell}$ in center $\ell$. Now, based on similar calculations as in the homogeneous setting
\begin{align*}
\sqrt{n}\big(\widehat{\btheta}_{1a,\BFI} - \btheta_{1a}\big) \;\leadsto\; {\mathcal N}\Big(0,\Big(\sum_{\ell=1}^L w_\ell \big(I_{1a,\ell}-I_{1ab,\ell}(I_{1b,\ell})^{-1}(I_{1ab,\ell})^t\big)\Big)^{-1}\Big),
\end{align*}
with the asymptotic covariance matrix equal to the asymptotic covariance matrix of the MAP estimator based on all data, given in Equation (\ref{Eq: inv cov}). The BFI estimator $\widehat{\btheta}_{1a,\BFI}$ is asymptotically efficient for estimating $\btheta_{1a}$. 

If $I_{1a,\ell}, I_{1b,\ell}$ and $I_{1ab,\ell}$ equal across the centers:
\begin{align*}
\Big(\sum_{\ell=1}^L w_\ell \big\{I_{1a,\ell}-I_{1ab,\ell}(I_{1b,\ell})^{-1}(I_{1ab,\ell})^t\big\}\Big)^{-1} = \Big(I_{1a,\ell}-I_{1ab,\ell}(I_{1b,\ell})^{-1}(I_{1ab,\ell})^t\Big)^{-1}.    
\end{align*}

\bigskip

\noindent
{\bf Asymptotic behaviour of the BFI estimator of $\btheta_{1b,\ell}$ }\\
Under the assumption that $\bLambda_{1b\ell}=\bLambda_{1b,\ell}$ and $w_\ell>0$, the BFI estimator for $\btheta_{1b,\ell}$ is given by
\begin{align*}
\widehat{\btheta}_{1b,\ell,\BFI} 
\;=\; \big(\widehat{\bA}_{1b,\ell}\big)^{-1}\Big[\widehat{\bA}_{1b,\ell}\widehat{\btheta}_{1b,\ell}
 + \big(\widehat{\bA}_{1ab,\ell}\big)^t\big(\widehat{\btheta}_{1a,\ell} 
 -
 \widehat{\btheta}_{1a,\BFI}\big)\Big] \;=\; \widehat{\btheta}_{1b,\ell}
 + \big(\widehat{\bA}_{1b,\ell}\big)^{-1}\big(\widehat{\bA}_{1ab,\ell}\big)^t\big(\widehat{\btheta}_{1a,\ell}  -  \widehat{\btheta}_{1a,\BFI}\big). 
 \end{align*}
Then,
\begin{align}
\sqrt{n}\big(\widehat{\btheta}_{1b,\ell,\BFI}-\btheta_{1b,\ell}\big) &=\; \sqrt{n}\big(\widehat{\btheta}_{1b,\ell}-\btheta_{1b,\ell}\big) + \big(\widehat{\bA}_{1b,\ell}\big)^{-1}\big(\widehat{\bA}_{1ab,\ell}\big)^t\sqrt{n}\big(\widehat{\btheta}_{1a,\ell}  -  \widehat{\btheta}_{1a,\BFI}\big)\nonumber \\[4pt] 
&=\; \sqrt{n}(\widehat{\btheta}_{1b,\ell}-\btheta_{1b,\ell}) + \big(\widehat{\bA}_{1b,\ell}\big)^{-1}\big(\widehat{\bA}_{1ab,\ell}\big)^t\sqrt{n}\big(\widehat{\btheta}_{1a,\ell}  -  \btheta_{1a}\big) \;+\; \big(\widehat{\bA}_{1b,\ell}\big)^{-1}\big(\widehat{\bA}_{1ab,\ell}\big)^t\sqrt{n}\big(\btheta_{1a}  -  \widehat{\btheta}_{1a,\BFI}\big)
\label{eq: theta1b}
\end{align}
We first leave out the last term and consider the asymptotic behaviour of 
\begin{align*}
\sqrt{n}\big(\widehat{\btheta}_{1b,\ell}-\btheta_{1b,\ell}\big) + \big(\widehat{\bA}_{1b,\ell}\big)^{-1}\big(\widehat{\bA}_{1ab,\ell}\big)^t\sqrt{n}\big(\widehat{\btheta}_{1a,\ell}  -  \btheta_{1a}\big).  
\end{align*}
As explained before, this term equals 
\begin{align*}
\sqrt{n}\big(\widehat{\btheta}_{1b,\ell}-\btheta_{1b,\ell}\big) &+ \big(I_{1b,\ell}\big)^{-1}\big(I_{1ab,\ell}\big)^t\sqrt{n}\big(\widehat{\btheta}_{1a,\ell}  -  \btheta_{1a}\big) \;+\; o_p(1)\\[4pt]
&=\; w_\ell^{-1/2} \sqrt{n_\ell}\big(\widehat{\btheta}_{1b,\ell}-\btheta_{1b,\ell}\big) + w_\ell^{-1/2}\big(I_{1b,\ell}\big)^{-1}\big(I_{1ab,\ell}\big)^t\sqrt{n_\ell}\big(\widehat{\btheta}_{1a,\ell}  -  \btheta_{1a}\big) \;+\; o_p(1).  
\end{align*}
In center $\ell$ the MAP estimator $\widehat{\btheta}_{1,\ell}$ is asymptotically normal (Bernstein-Von Mises Theorem\cite{vdVaart}):
\begin{align*}
\sqrt{n_\ell} \begin{pmatrix} \begin{pmatrix}
\widehat{\btheta}_{1a,\ell}\\
\widehat{\btheta}_{1b,\ell}
\end{pmatrix}
-
\begin{pmatrix}
\btheta_{1a}\\
\btheta_{1b,\ell}
\end{pmatrix}
\end{pmatrix} \;\leadsto \; {\mathcal N}\big({\bf 0}, (I_{1,\ell})^{-1} \big),
\end{align*}
Define the function $g(\btheta_{1a},\btheta_{1b,\ell}) = \btheta_{1b,\ell} + \big(I_{1b,\ell}\big)^{-1}\big(I_{1ab,\ell}\big)^t \btheta_{1a}$. Then, by the continuous mapping theorem
\begin{align*}
\sqrt{n_\ell}(\widehat{\btheta}_{1b,\ell}-\btheta_{1b,\ell}) &+ \big(I_{1b,\ell}\big)^{-1}\big(I_{1ab,\ell}\big)^t\sqrt{n_\ell}\big(\widehat{\btheta}_{1a,\ell}  -  \btheta_{1a}\big) =\; \sqrt{n}_\ell \begin{pmatrix} g\begin{pmatrix}
\widehat{\btheta}_{1a,\ell}\\
\widehat{\btheta}_{1b,\ell}
\end{pmatrix}
-
g\begin{pmatrix}
\btheta_{1a}\\
\btheta_{1b,\ell}
\end{pmatrix}
\end{pmatrix}\\[8pt] 
&\leadsto \; {\mathcal N}\big({\bf 0}, g'(\btheta_{1a},\btheta_{1b,\ell})^t \big(I_{1,\ell}\big)^{-1} g'(\btheta_{1a},\btheta_{1b,\ell}) \big) 
\end{align*}
with $g'(\btheta_{1a},\btheta_{1b,\ell})$ the derivative of the function $g$ in $(\btheta_{1a},\btheta_{1b,\ell})$. 
%= \big(\big(I_{1b,\ell}\big)^{-1}\big(I_{1ab,\ell}\big)^t, {\bf 1})$, with ${\bf 1}$ an identity matrix. 
Straightforward calculations show that $$g'(\btheta_{1a},\btheta_{1b,\ell})^t \big(I_{1,\ell}\big)^{-1} g'(\btheta_{1a},\btheta_{1b,\ell}) = (I_{1b,\ell})^{-1},$$ which
equals the asymptotic covariance matrix of the Gaussian limit distribution for $\widehat{\btheta}_{1b,\ell}$ if the parameter $\btheta_{1a}$ is known. Apparently, leaving out the last term in equation (\ref{eq: theta1b}), means that we assume that the BFI estimator for $\btheta_{1a}$ is (almost) equal to the true value $\btheta_{1a}$, and thus that $\btheta_{1a}$ is (almost) known. The result is interesting; the asymptotic accuracy of the BFI estimator for $\btheta_{1b,\ell}$ is increased, because the parameter $\btheta_{1a}$ can be estimated more accurately with the BFI estimator (and thus using information from the other centers) compared to the situation in which $\btheta_{1a}$ is estimated based on data from center $\ell$ only.      

We go back to the expression in (\ref{eq: theta1b}):
\begin{align*}
\sqrt{n}&(\widehat{\btheta}_{1b,\ell,\BFI}-\btheta_{1b,\ell}) 
\;=\; \sqrt{n}(\widehat{\btheta}_{1b,\ell}-\btheta_{1b,\ell}) + \big(\widehat{\bA}_{1b,\ell}\big)^{-1}\big(\widehat{\bA}_{1ab,\ell}\big)^t\sqrt{n}\big(\widehat{\btheta}_{1a,\ell}  - \widehat{\btheta}_{1a,\BFI}\big)\\[4pt]
%&=\; \sqrt{n_\ell}(\widehat{\btheta}_{1b,\ell}-\btheta_{1b,\ell}) + \big(I_{1b,\ell}\big)^{-1}\big(I_{1ab,\ell}\big)^t\sqrt{n_\ell}\big(\widehat{\btheta}_{1a,\ell}  -  \widehat{\btheta}_{1a}\big) + o_P(1)\\[8pt]
%&=\; \sqrt{n_\ell}(\widehat{\btheta}_{1b,\ell}-\btheta_{1b,\ell}) - \big(I_{1b,\ell}\big)^{-1}\big(I_{1ab,\ell}\big)^t\sqrt{n_\ell}\Bigg(\Big(\sum_{k=1}^L \frac{n_k}{n}\;\widehat{\Gamma}_k + \frac{1}{n}\;\bLambda_{1a}-\sum_{k=1}^L\frac{1}{n}\;\bLambda_{1a,k}  \Big)^{\!-1}  \sum_{k=1}^L \frac{n_k}{n}\;\widehat{\Gamma}_k \;\widehat{\btheta}_{1a,k} - \widehat{\btheta}_{1a,\ell}\Bigg) + o_P(1)\\[8pt]
&=\; \sqrt{n}(\widehat{\btheta}_{1b,\ell}-\btheta_{1b,\ell}) - \big(I_{1b,\ell}\big)^{-1}\big(I_{1ab,\ell}\big)^t\Bigg(\Big(\sum_{k=1}^L \frac{n_k}{n}\;\widehat{\Gamma}_k  \Big)^{\!-1}  \sum_{k=1}^L \frac{n_k}{n}\;\widehat{\Gamma}_k \;\sqrt{n}\big(\widehat{\btheta}_{1a,k} - \widehat{\btheta}_{1a,\ell}\big)\Bigg) + o_P(1)\\[8pt]
%&=\; \sqrt{n_\ell}(\widehat{\btheta}_{1b,\ell}-\btheta_{1b,\ell}) - \big(I_{1b,\ell}\big)^{-1}\big(I_{1ab,\ell}\big)^t\Bigg(\Big(\sum_{k=1}^L w_k\;\widehat{\Gamma}_k  \Big)^{\!-1}  \sum_{k=1}^L w_k\;\widehat{\Gamma}_k \;\sqrt{n_\ell}\big(\widehat{\btheta}_{1a,k} - \widehat{\btheta}_{1a,\ell}\big)\Bigg) + o_P(1)\\
&=\; \sqrt{n}(\widehat{\btheta}_{1b,\ell}-\btheta_{1b,\ell}) - \big(I_{1b,\ell}\big)^{-1}\big(I_{1ab,\ell}\big)^t\Bigg(\Big(\sum_{k=1}^L w_k\;\Gamma_k  \Big)^{\!-1}  \sum_{k=1,k\neq \ell}^L w_k\;\Gamma_k \;\big(\sqrt{n}(\widehat{\btheta}_{1a,k} - \btheta_{1a}) - \sqrt{n}\big(\widehat{\btheta}_{1a,\ell} - \btheta_{1a})  \big)\Bigg) + o_P(1)\\
&=\; \sqrt{n}(\widehat{\btheta}_{1b,\ell}-\btheta_{1b,\ell}) + \sqrt{n}\big(\widehat{\btheta}_{1a,\ell} - \btheta_{1a}\big) \big(I_{1b,\ell}\big)^{-1}\big(I_{1ab,\ell}\big)^t\Big(\sum_{k=1}^L w_k\;\Gamma_k  \Big)^{\!-1}  \sum_{k=1,k\neq \ell}^L w_k\;\Gamma_k\\
& ~~~~ -\big(I_{1b,\ell}\big)^{-1}\big(I_{1ab,\ell}\big)^t\Big(\sum_{k=1}^L w_k\;\Gamma_k  \Big)^{\!-1}  \sum_{k=1,k\neq \ell}^L w_k\;\Gamma_k \;\sqrt{n}\big(\widehat{\btheta}_{1a,k} - \btheta_{1a}\big) \; + o_P(1).
\end{align*}
The asymptotic distribution can be obtained with the continuous mapping theorem and Slutky's lemma again. The third and last term  depends on data from all centers except from center $\ell$. Since the data from the different centers are assumed to be independent, it is sufficient to show that the asymptotic distributions of the sum of the first and second terms and of the third term are asymptotically normal and next add the asymptotic mean (which are zero) and the variances. The asymptotic distribution of the sum of the first and second term can obtained with the continuous mapping theorem, like before. The third term is asymptotically normal:  
\begin{align*}
\big(I_{1b,\ell}\big)^{-1}\big(I_{1ab,\ell}\big)^t\Big(\sum_{k=1}^L w_k\;\Gamma_k  \Big)^{\!-1}  \sum_{k=1,k\neq \ell}^L \sqrt{w_k}\;\Gamma_k \;\sqrt{n_k}\big(\widehat{\btheta}_{1a,k} - \btheta_{1a}\big) \leadsto {\mathcal N}({\bf 0}, \Sigma)    
\end{align*}
with
\begin{align*}
\Sigma &= \big(I_{1b,\ell}\big)^{-1} \big(I_{1ab,\ell}\big)^{t} \Big(\sum_{k=1}^L w_k \Gamma_k\Big)^{-1}\Big(\sum_{k\neq l}w_k\Gamma_k\Big)\Big(\sum_{k=1}^L w_k \Gamma_k\Big)^{-1} I_{1ab,\ell} \big(I_{1b,\ell}\big)^{-1} \\
&= \big(I_{1b,\ell}\big)^{-1} \big(I_{1ab,\ell}\big)^{t} \Big(\sum_{k=1}^L w_k \Gamma_k\Big)^{-1}\Big(\sum_{k=1}^Lw_k\Gamma_k - w_\ell\Gamma_\ell\Big)\Big(\sum_{k=1}^L w_k \Gamma_k\Big)^{-1} I_{1ab,\ell} \big(I_{1b,\ell}\big)^{-1}\\
&= \big(I_{1b,\ell}\big)^{-1} \big(I_{1ab,\ell}\big)^{t} \Big(I - \Big(\sum_{k=1}^L w_k \Gamma_k\Big)^{-1}w_\ell\Gamma_\ell\Big)\Big(\sum_{k=1}^L w_k \Gamma_k\Big)^{-1} I_{1ab,\ell} \big(I_{1b,\ell}\big)^{-1}\\
&= \big(I_{1b,\ell}\big)^{-1} \big(I_{1ab,\ell}\big)^{t} \Big(\sum_{k=1}^L w_k \Gamma_k\Big)^{-1} I_{1ab,\ell} \big(I_{1b,\ell}\big)^{-1} - \big(I_{1b,\ell}\big)^{-1} \big(I_{1ab,\ell}\big)^{t} \Big(\sum_{k=1}^L w_k \Gamma_k\Big)^{-1}w_\ell\Gamma_\ell\Big(\sum_{k=1}^L w_k \Gamma_k\Big)^{-1} I_{1ab,\ell} \big(I_{1b,\ell}\big)^{-1}.
\end{align*}

If we sum up the variances of the asymptotic zero mean Gaussian distributions of all terms, we obtain
\begin{align*}
\big(I_{1b,\ell}\big)^{-1} \big(I_{1ab,\ell}\big)^{t} \Big(\sum_{k=1}^L w_k \Gamma_k\Big)^{-1} I_{1ab,\ell} \big(I_{1b,\ell}\big)^{-1} + \big(w_\ell I_{1b,\ell}\big)^{-1}
\end{align*}
So, 
\begin{align*}
\sqrt{n}\big(\widehat{\btheta}_{1b,\ell,\BFI} - \btheta_{1b,\ell}\big) \leadsto {\mathcal N}\Big({\bf 0}, \big(w_\ell I_{1b,\ell}\big)^{-1}+\big(I_{1b,\ell}\big)^{-1} \big(I_{1ab,\ell}\big)^{t} \Big(\sum_{k=1}^L w_k \big(I_{1a,k}-I_{1ab,k}(I_{1b,k})^{-1}(I_{1ab,k})^t\big)\Big)^{-1} I_{1ab,\ell} \big(I_{1b,\ell}\big)^{-1}\Big). 
\end{align*}
The BFI estimator follows, asymptotically, the same distribution as the MAP estimator based on the combined data, see Equation (\ref{eq: asympt 1b}), and is asymptotically efficient. If $\bLambda_{1b\ell} \neq \bLambda_{1b,\ell}$ the asymptotic distribution of $\widehat{\btheta}_{1b,\ell,\BFI}$ will not change, because $n^{-1}\bLambda_{1b\ell}$ and $n^{-1}\bLambda_{1b,\ell}$ converge to zero. 

Like in the homogeneous setting, it can be directly seen that the BFI estimator $n^{-1} \widehat{\bA}_{1a,\BFI}$ converges in probability to $\sum_{\ell=1}^L w_\ell I_{1a,\ell}$.
Similarly, the BFI estimator $n^{-1} \widehat{\bA}_{1b,\ell,\BFI}$ converges in probability to $w_\ell I_{1b,\ell}$, and $n^{-1} \widehat{\bA}_{1ab,\ell,\BFI}$ to $w_\ell I_{1ab,\ell}$. %Computing the inverse of the matrix $I_1$ that consists of the different blocks like $J_1$ does, yields the asymptotic covariance matrices for $\widehat{\btheta}_{1a,\BFI}$ and $\widehat{\btheta}_{1b,\ell,\BFI}$.   

\subsubsection*{Asymptotic distribution of the WAV and single center estimators}
If the parameter $\btheta_{1a}$ is estimated by the MAP estimator from center $\ell$, $\widehat{\btheta}_{1a,\ell}$, the asymptotic distribution equals
\begin{align*}
\sqrt{n}\big(\widehat{\btheta}_{1a,\ell} - \btheta_{1a}\big) \;\leadsto\; {\mathcal N}\Big(0,w_\ell^{-1}\Big(I_{1a,\ell}-I_{1ab,\ell}(I_{1b,\ell})^{-1}(I_{1ab,\ell})^t\Big)^{-1}\Big).
\end{align*}
If the weighted average estimator $\sum_{\ell=1}^L \frac{n_\ell}{n} \widehat{\btheta}_{1a,\ell}$ is used, the asymptotic distribution equals, like before
\begin{align*}
\sqrt{n}\Big(\sum_{\ell=1}^L \frac{n_\ell}{n}\; \widehat{\btheta}_{1a,\ell} - \btheta_{1a}\Big) \;\leadsto\; {\mathcal N}\Big(0,\sum_{\ell=1}^Lw_\ell\Big( I_{1a,\ell}-I_{1ab,\ell}(I_{1b,\ell})^{-1}(I_{1ab,\ell})^t\Big)^{-1}\Big).
\end{align*}
Asymptotically, both estimators are not efficient,  unless $I_{1,\ell}=I_1, \ell=1,\ldots,L$. In that case the weighted average estimator is asymptotically efficient as well.

If we estimate $\btheta_{1b,\ell}$ by its MAP estimator in the corresponding center, the asymptotic distribution equals:
\begin{align*}
\sqrt{n}\big(\widehat{\btheta}_{1b,\ell} - \btheta_{1b,\ell}\big) \leadsto {\mathcal N}\Bigg({\bf 0}, w_\ell^{-1}\Big( I_{1b,\ell}^{-1} + \big(I_{1b,\ell}\big)^{-1} (I_{1ab,\ell})^t \big(I_{1a,\ell}-I_{1ab,\ell} (I_{1b,\ell})^{-1} (I_{1ab,\ell})^t\big)^{-1} I_{1ab,\ell} (I_{1b,\ell})^{-1}\Big)\Bigg).
\end{align*}
Since $\btheta_{1b,\ell}$ is center-specific, the WAV estimator equals the single center estimator. The estimator $(\widehat{\btheta}_{1a,\ell},\widehat{\btheta}_{1b,\ell})$ is based on data from center $\ell$ only, and does not use any information from the other centers for estimating the parameter $\btheta_{1a}$. Using information from the other centers, $\btheta_{1a}$ is estimated more accurately and improves the estimation of $\btheta_{1b}$. That is why the BFI estimator $\widehat{\btheta}_{1b,\ell,\BFI}$ is more accurate than the weighted average estimator.  

\bigskip

The asymptotic distribution of the BFI estimators for clustered data (Appendix II.C) can be derived in a similar way, but this is more complicated due to the complex expressions of these estimators. However, since the BFI estimators in the homogeneous and the heterogeneous case with center-specific parameters have shown to be asymptotically efficient and these settings are special cases of the one with clustered data, it is expected that the BFI estimators for the clustered data are asymptotically efficient as well.

\end{document}